\documentclass{article}

\usepackage{arxiv}

\usepackage[utf8]{inputenc} 
\usepackage[T1]{fontenc}    
\usepackage{hyperref}       
\usepackage{url}            
\usepackage{booktabs}       
\usepackage{amsfonts}       
\usepackage{nicefrac}       
\usepackage{microtype}      
\usepackage[title]{appendix}
\usepackage{graphicx}
\usepackage[bf]{caption}

\usepackage{doi}
\usepackage{etoolbox}
\usepackage[natbibapa]{apacite}
\AtBeginDocument{}
\renewcommand{\APACrefnote}[1]{}

\newtoggle{bibdoi}
\newtoggle{biburl}
\makeatletter
\newsavebox{\bib@url}
\newsavebox{\bib@doi}

\undef{\APACrefURL}
\undef{\endAPACrefURL}
\undef{\APACrefDOI}
\undef{\endAPACrefDOI}

\newenvironment{APACrefURL}
  {\global\toggletrue{biburl}\lrbox\bib@url}
  {\endlrbox}

\newenvironment{APACrefDOI}
  {\global\toggletrue{bibdoi}\lrbox\bib@doi}
  {\endlrbox}

\newcommand{\printinfo}{
  \iftoggle{bibdoi}{\usebox{\bib@doi}}{\usebox{\bib@url}}
  \togglefalse{bibdoi}
}

\AtBeginEnvironment{thebibliography}{
\pretocmd{\PrintBackRefs}{%
  \iftoggle{bibdoi}
    {\iftoggle{biburl}{\unskip\unskip}{}\usebox{\bib@doi}}
    {\iftoggle{biburl}{Retrieved from \usebox{\bib@url}}}{}
  \togglefalse{bibdoi}\togglefalse{biburl}%
}{}{}}
\makeatother

\title{Quantitative Measures for Integrating Resilience into Transportation Planning Practice: Study in Texas}

\date{} 					

\renewcommand{\headeright}{}
\renewcommand{\undertitle}{}
\renewcommand{\shorttitle}{}

\hypersetup{
pdftitle={Quantitative Measures for Integrating Resilience into Transportation Planning Practice: Study in Texas},
pdfsubject={},
}

\begin{document}
\maketitle

\begin{center}
{\Large
Cheng-Chun Lee\textsuperscript{a,*},
Akhil Anil Rajput\textsuperscript{a}, 
Chia-Wei Hsu\textsuperscript{a}, 
Chao Fan\textsuperscript{a}, 
Faxi Yuan\textsuperscript{a}, 
Shangjia Dong\textsuperscript{b}, 
Amir Esmalian\textsuperscript{a}, 
Hamed Farahmand\textsuperscript{a}, 
Flavia Ioana Patrascu\textsuperscript{a}, 
Chia-Fu Liu\textsuperscript{a}, 
Bo Li\textsuperscript{a}, 
Junwei Ma\textsuperscript{a}, 
Ali Mostafavi\textsuperscript{a}
\par}

\bigskip
\textsuperscript{a} Urban Resilience.AI Lab, Zachry Department of Civil and Environmental Engineering,\\ Texas A\&M University, 199 Spence St., College Station, TX 77843\\
\vspace{6pt}
\textsuperscript{b} Department of Civil and Environmental Engineering,\\ University of Delaware, Newark, United States\\
\vspace{6pt}
\textsuperscript{*} correseponding author, email: ccbarrylee@tamu.edu
\\
\end{center}
\bigskip
\begin{abstract}
The objective of this study is to propose a system-level framework with quantitative measures to assess the resilience of road networks. The framework proposed in this paper can help transportation agencies incorporate resilience considerations into project development proactively and to understand the resilience performance of current road networks effectively. This study identified and implemented four quantitative metrics to classify the criticality of road segments based on critical dimensions of road network resilience, and two integrated metrics were proposed to combine all metrics to show the overall resilience performance of road segments. A case study was conducted on the Texas road networks to demonstrate the effectiveness of implementing this framework in a practical scenario. Since the data used in this study is available to other states and countries, the framework presented in this study can be adopted by other transportation agencies across the globe for regional transportation resilience assessments.
\end{abstract}



\section{Introduction}
The objective of the study is to propose a system-level framework with quantitative measures to assess the resilience of road networks to inform transportation planning and project development processes in practice. Road networks are the backbone of society that enables the movement of people and goods safely and efficiently. A reliable road network does not only allow improved access by the workforce to business, but also result in long-term cost saving and increase of safety and viability of residential neighborhoods \citep{weilant_incorporating_2019}. Disruptions of road networks would reduce economic productivity, harm local commercial activities and community well-being, and restrict people’s mobility and access to critical facilities \citep{weilant_incorporating_2019,frizzelle_importance_2009,jenelius_importance_2006}. The US National Response Plan (NRP) urges each state to have a plan for responding to “… reduce the vulnerability to all natural and manmade hazards; and minimize the damage and assist in the recovery from any type of incident that occurs”  \citep{us_department_of_homeland_security_national_2004}. The need for incorporating resilience assessments in transportation planning and project development processes is increasingly recognized by transportation organizations in the United States and worldwide. For example, the US Federal Highway Administration (FHWA) has established mandates and guidelines for transportation agencies regarding the incorporation of resilience into transportation systems through their asset management planning efforts \citep{flannery_resilience_2018}.

While the need for system-level resilience assessment has been recognized, the measures for quantifying system-level resilience to inform transportation planning processes are rather limited in practice \citep{esmalian_operationalizing_2022}. Existing literature has shown that the evaluation of road network resilience is multi-dimensional, including vulnerability assessment, criticality assessment, and interdependency and cascading failure assessment. \citet{sharma_resilience_2018} suggested measuring resilience using vulnerability, severity of consequences, and time to recover from the disruptions. Considering the post-disaster network access to critical facilities, studies \citep{dong_robust_2019, dong_integrated_2020} proposed the concept of robust components to measure communities’ risk disparity in terms of losing access to hospitals in both earthquake and flooding scenarios using a network topology analysis approach. \citet{gao_universal_2016} used a single metric to measure the resilience of multi-dimensional complex networks. The author found that a resilient system is (1) densely connected in the system, (2) highly heterogeneous in the connections, and (3) relatively symmetrical. Using a percolation approach, \citet{dong_measuring_2020} measured the resilience of different cities’ and states’ road networks in the face of random failure using the giant component metrics. This topological metric enables the ranking of road networks based on the derived resilience score. \citet{weilant_incorporating_2019} summarized the resilience measurement process into an AREA approach, which represents Absorptive capacity, Restorative capacity, Equitable access, and Adaptive capacity. \citet{bruneau_framework_2003} proposed that a resilience framework should include measurement of robustness, rapidity, resourcefulness, and redundancy, integrated with technical, organizational, social, and economic dimensions \citep{davis_establishing_2018, flannery_resilience_2018, norris_community_2008, reggiani_network_2013}. \citet{da_silva_city_2014} suggested a resilience system should be equipped with qualities such as robustness (can anticipate potential failures and make provisions), redundancy (has spare capacity to accommodate changes), and integratedness (promotes consistency by pursuing integration and alignment between systems). This study explored various quantitative metrics and identified those in this multi-dimensional and system-level assessment framework based on connectivity of road segments in road networks; vulnerability of road segments to extreme events, such as flooding and hurricanes; disrupted access to critical facilities; and cascading impact of critical facility networks. Quantitative metrics that can consider all of these dimensions in resilience assessment are essential for a system-level understanding of the resilience of road networks and also inform transportation planning and project development.

While the scholarly literature is rich in terms of methods and measures for assessment of vulnerability and resilience in transportation infrastructure, little of the existing work has been adopted in practice. The literature has put efforts into developing assessment measures to understand the resilience and vulnerability of road networks \citep{hsieh_highway_2020,singh_vulnerability_2018,dave_extreme_2021,martin_assessing_2021}. Several conceptual frameworks have been proposed in the literature to promote consideration of resilience into infrastructure assessment, including road networks \citep{jenelius_road_2015,morelli_measuring_2021,papilloud_vulnerability_2021}. However, in the United States, only about half of the state departments of transportation (DOTs) and metropolitan planning organizations (MPOs) started incorporating resilience considerations in their plans, and only some of them explicitly listed resilience as a goal \citep{dix_integrating_2018}. This limitation is mainly due to two reasons: (1) transportation planners and decision-makers do not know what method and measures to use among several proposed in the literature; (2) the methods and measures for vulnerability and resilience assessment should be obtainable using the existing data available to most transportation agencies. 

\begin{figure}
	\centering
    \includegraphics[width=0.6\linewidth]{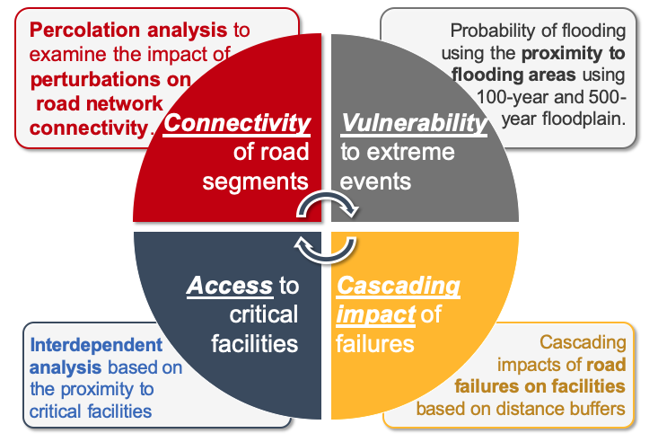}
    \caption{Four metrics for understanding the resilience of road networks. The quantitative and systematic evaluation metrics are informed by the existing literature.}
	\label{fig:fig1}
\end{figure}

To address this gap, this study identified and implemented a framework incorporating quantitative metrics to classify the criticality of road segments for understanding the system-level vulnerability and criticality of the road network of the state of Texas to inform the planning and project development processes. As shown in Figure \ref{fig:fig1}, the proposed framework includes four quantitative metrics that quantify vulnerability and criticality for each segment of the state road network. The first metric for connectivity of road segments adopts percolation analysis to quantify the criticality of road segments under the impact of external perturbations on road networks. The second metric for vulnerability to extreme events examines the proximity of road segments to the disaster impact areas. In this study, we identified areas exposed for flood hazards using 100- and 500-year floodplains to quantify the probability of flooding on road segments. The third and fourth metrics are based on interdependency analysis and proximity to critical facilities and infrastructure networks to quantify the criticality of road segments for accessing critical facilities. Finally, we combined the four metrics into integrated metrics to obtain the overall criticality of the road segments in road networks. In addition, this study presents a case study by applying the proposed framework in Texas road networks to demonstrate the effectiveness of implementing this framework in a practical scenario.

The remainder of the paper is structured as follows: Section \ref{sec:Research Approach} outlines the research approach of the four quantitative metrics used for the resilience assessment of road networks. Sections \ref{sec:Connectivity of Road Segments} to Section \ref{sec:Cascading Impact of Critical Infrastructure Networks} provide detailed information and discussion of each quantitative metric. Section \ref{sec:Integrated Criticality Metric} demonstrates details of integrated metrics that consider all individual metrics in different aspects. Section \ref{sec:Case Study for Texas Road Network} presents a case study in Texas road networks by implementing the proposed framework. Finally, Section \ref{sec:Concluding Remarks} presents conclusions gleaned from the study.

\section{Research Approach}
\label{sec:Research Approach}
\begin{figure}
	\centering
    \includegraphics[width=0.95\linewidth]{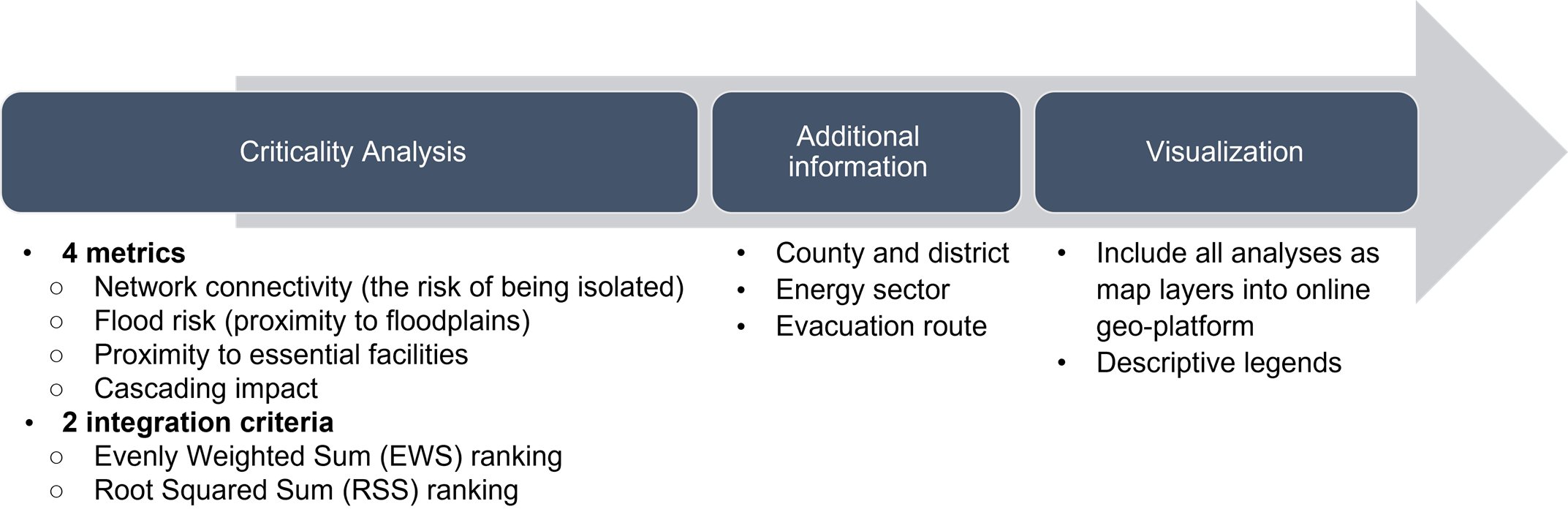}
    \caption{Research approach for conducting the vulnerability and resilience assessment of state road infrastructure networks.}
	\label{fig:fig2}
\end{figure}

This study implemented quantitative and computational research methods for the resilience assessment of road networks. As shown in Figure \ref{fig:fig2}, the research approach first identified quantitative metrics for the resilience assessment of road networks. With those metrics, we then incorporated additional information, such as evacuation routes, to understand the criticality of road networks. Lastly, all of the results and information were integrated into a geographic information system-based map for practical usage. In this section, we briefly outline how the quantitative metrics are calculated for road networks; further specific discussions can be found in the sections devoted to these metrics.

First, we evaluate the criticality of individual road segments for the connectivity of the road network, a characteristic that is particularly important in disasters. For example, residents rely on connected roads to evacuate, and first responders need to access impacted areas to save lives and protect properties. During disasters, fragmented networks may isolate areas, block transportation resources, and create traffic congestion. In addition, disruptions or floodwaters on road networks may spread over time. Measures considering static road network failures only may fail to capture the interconnections of road segments and consequent impacts of flood propagation. Hence, in this task, we adopted percolation analysis to simulate how road failures spread from one road segment to another and the extent to which failures of road networks influence their connectivity. In the case study, we collected road network data from the Texas Department of Transportation (TxDOT) roadway inventory On-System dataset updated on August 8, 2021, to perform these analyses and compute this criticality metric for road segments. This dataset includes details about the geometries of the road segments, road types, speed limits, and names of the roads. Also, the data includes both highways and streets in urban and rural areas in Texas.

The second metric is the criticality of road segments to disaster impacts, which, in this study, are caused by flooding. The area of the case study, the state of Texas, especially South Texas, is prone to extreme events, such as flooding and hurricanes. Determining the criticality of road segments during flooding events is important for planning and prioritization in response to disruptive impacts of flooding. Hence, we collected 100- and 500-year floodplain data, which is the most recent nationwide extract of the National Flood Hazard Layer (NFHL) for Texas. The NFHL dataset was obtained by the Federal Emergency Management Agency (FEMA) for examining road exposure to flooding. The data includes flood map panel boundaries, flood hazard zone boundaries, and other information related to flood control zones and areas. Using this data, we identified the areas of flood zones and calculated the proximity of road networks to the nearest flood zones. Based on the proximity, we classified the distance into five categories representing the level of criticality of the road segments to flood.

Third, we analyzed the criticality of road segments based on impacts of road disruption on access to (1) critical facilities such as hospitals and fire stations and (2) critical infrastructure networks such as gas lines and oil pipelines. One essential role of the road networks is to transport resources and provide public services. Accessing critical facilities is thus an important component of road network resilience assessment. We implemented two metrics to evaluate the criticality of road segments based on the interdependencies, measured based on distance buffers, between road segments and facilities. In particular, one metric is measured using the proximity of road segments to criticality facilities, while another metric is measured based on the closeness to critical infrastructure networks. We collected facility location data from Homeland Infrastructure Foundation Level Data (HIFLD) in the ArcGIS data repository, as well as energy sector data and evacuation route data from TxDOT, which include information of facility types and corresponding locations. We then extracted the location information for each of the critical facilities, such as electrical substations; power plants; hospitals; national shelter facilities; seaports; petroleum, oil and lubricant (POL) terminals; fire stations; and police stations, as well as for each of the critical infrastructure networks, such as railroad network, transmission lines, gas lines, hydrocarbon gas lines, and oil pipelines, which are essential for resiliency. Disruption in access to these facilities and infrastructure networks can exacerbate the impacts faced by communities during and in the aftermath of disasters. 

Finally, it is important to evaluate the overall criticality by considering all dimensions of the resilience of road networks. To this end, we devised two integration approaches to show the overall criticality of the road segments. The first method is to rank overall criticality based on the evenly weighted sum of all individual criticality metrics of each road segment. The second approach is based on the root squared sum ranking, by which the overall criticality score is calculated by taking the square root of the sum of squares of individual metrics. In the case study, we computed the integrated criticality metrics for all road segments in Texas. The results for all metrics — including four individual and two integrated metrics — can be mapped onto GIS-based maps for visualization and to provide opportunities for further incorporation with the other resilience dimensions or critical considerations. Detailed information for each metric, including individual and integrated metrics, is discussed in the following sections.

\section{Connectivity of Road Segments}
\label{sec:Connectivity of Road Segments}
\subsection{Metric Definition and Calculation}
The connectivity of a road segment is determined by the percolation analysis performed on the road network. The percolation analysis is an analytical technique that determines to what extent a road segment contributes to the network’s overall connectivity. The procedure for percolation analysis is to remove nodes and links based on specific criteria and to quantify changes in network structure after a disturbance. When more critical road segments are perturbed, they are more likely to cause significant disconnection in the network.

\begin{figure}
	\centering
    \includegraphics[width=0.85\linewidth]{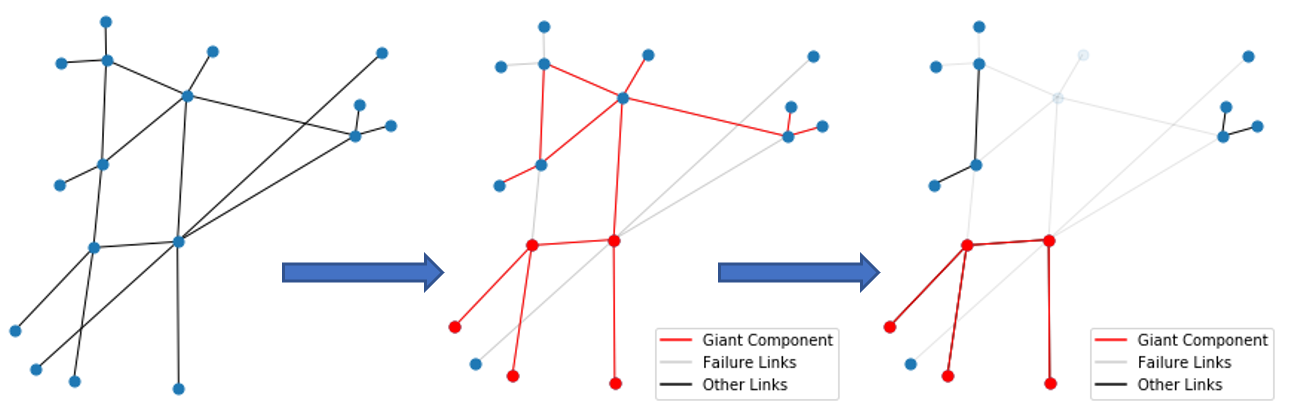}
    \caption{Example of percolation analysis by removing failure links and recording the size of the giant component.}
	\label{fig:fig3}
\end{figure}

\begin{figure}
	\centering
    \includegraphics[width=0.95\linewidth]{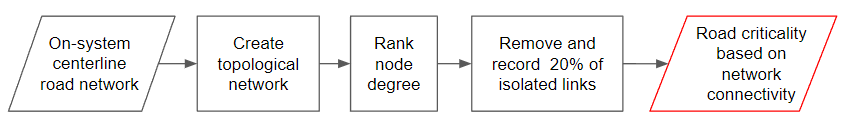}
    \caption{Steps for evaluating the criticality of road segments based on network connectivity.}
	\label{fig:fig4}
\end{figure}

While natural hazards, such as flooding, perturb road networks, some components within the networks may fail and other parts of the networks could be disconnected and isolated. The percolation analysis mimics this process by sequentially removing nodes in topological networks. Standard criteria of node attributes rank the criticality of components within a network: degree, weighted degree, link betweenness, and size of the giant component. Figure \ref{fig:fig3} shows an example of removing failure links and recording the size of giant components to understand the impact of perturbations on road networks. To obtain the worst scenario, we applied the node degree link removal strategy for simulation. That is, to identify failure links we removed nodes in topological networks in order from those with high degrees (node having a greater number of connections) to low degrees (node having fewer connections). In the real-world situation, it is unlikely that all the nodes with higher degrees would fail at the early stage simultaneously. During each removal stage in this analysis, essential characteristics were recorded, including the proportion of isolated links and critical transitions. Isolated links are those that become disconnected from any other part of a network; and critical transitions occur where network connectivity significantly drops (proportion of isolated links suddenly increase). Figure \ref{fig:fig4} depicts the steps for computing this metric. In this study, the road segments are grouped into five criticality levels. The first 20\% of isolated road segments are assigned level 1, and the last 20\%, level 5.

\subsection{Data Collection and Processing for the Case Study}
In the case study, we utilized roadway inventory data updated on August 8, 2021, as the road network for computing this metric. The original road network data contains multiple lanes and complicated intersections. To perform network analysis, we simplified the original road network data, including merging multiple lanes on the same road segment into a single central line if they are close enough to each other. The simplified road network is then transformed into a topological network where all intersections and road segments are represented as nodes and links.

\section{Vulnerability to Flooding}
\label{sec:Vulnerability to Flooding}
This metric promotes understanding of the criticality of road segments to disaster impacts: flooding, in the case study. According to the literature, flooding is the most prominent hazard of concern for the Texas DOT and Texas MPOs. Therefore, we included a metric focusing on the exposure to flood hazards determined by the proximity of road segments to floodplains. This metric can be adopted to account for other disaster impacts to evaluate the criticality of road segments in road networks.

\subsection{Metric Definition and Calculation}
\begin{figure}
	\centering
    \includegraphics[width=0.75\linewidth]{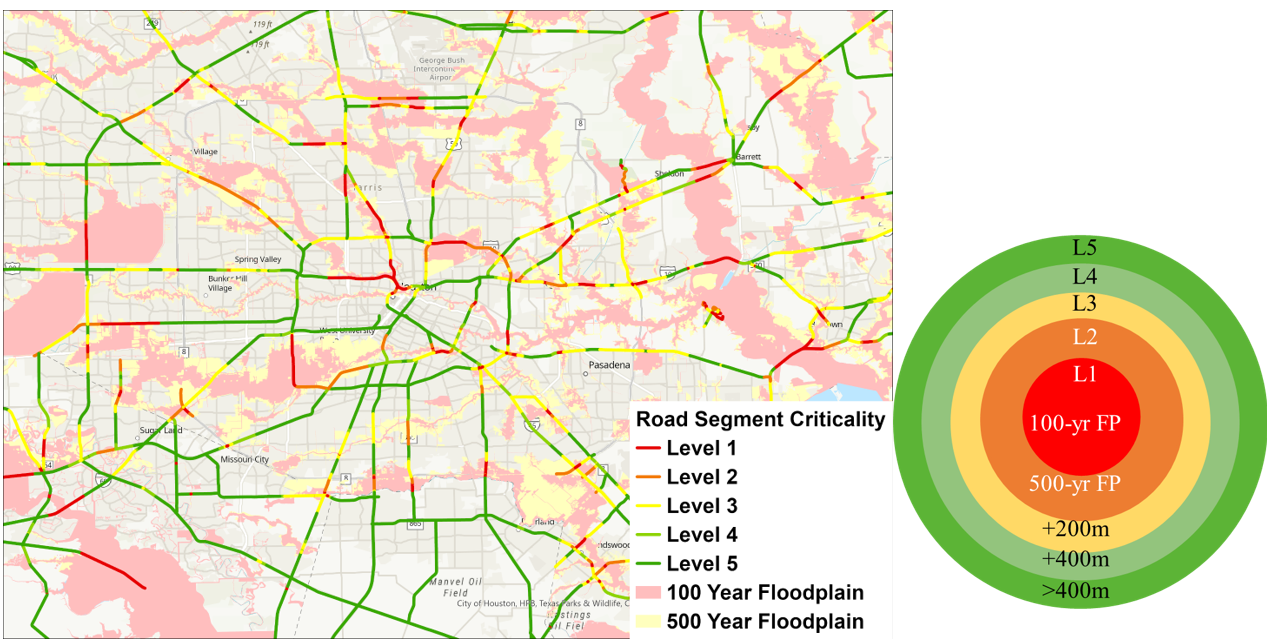}
    \caption{Definition of the metric of vulnerability to extreme events and an example of criticality levels of road segments in central Houston.}
	\label{fig:fig5}
\end{figure}

\begin{figure}
	\centering
    \includegraphics[width=0.95\linewidth]{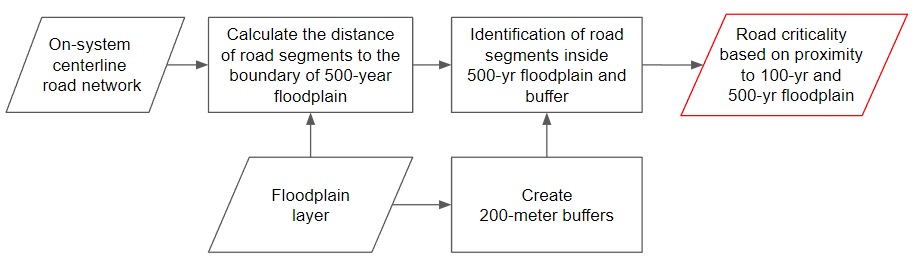}
    \caption{Steps for evaluating the criticality of road segments based on proximity to flooding areas.}
	\label{fig:fig6}
\end{figure}

This metric determines the criticality level of road segments based on their proximity to floodplains. ArcGIS was used to perform geoprocessing tasks, including: (1) distinct road segments whose midpoints are located in the 100-year floodplain and the 500-year floodplain, and (2) calculating the distances from the midpoint of the rest of the road segments to the boundary of the 500-year floodplain. As shown in Figure \ref{fig:fig5}, this metric determines the road segments belonging to one of the five criticality levels. The road segments with their midpoint located in the 100-year floodplain are assigned to level 1. The road segments with their midpoint located between 100- and 500-year floodplain are assigned to level 2. Level 3 covers the road segments whose midpoints are 0 to 200 meters away from the 500-year floodplain. Level 4 covers the road segments whose midpoints are 200 to 400 meters away from the 500-year floodplain. Level 5 covers the road segments more than 400 meters away from the 500-year floodplain. Figure \ref{fig:fig5} demonstrates an example of assigning criticality levels to road segments in central Houston. Figure \ref{fig:fig6} depicts the steps for computing this metric.

\subsection{Data Collection and Processing for the Case Study}
\begin{figure}
	\centering
    \includegraphics[width=0.7\linewidth]{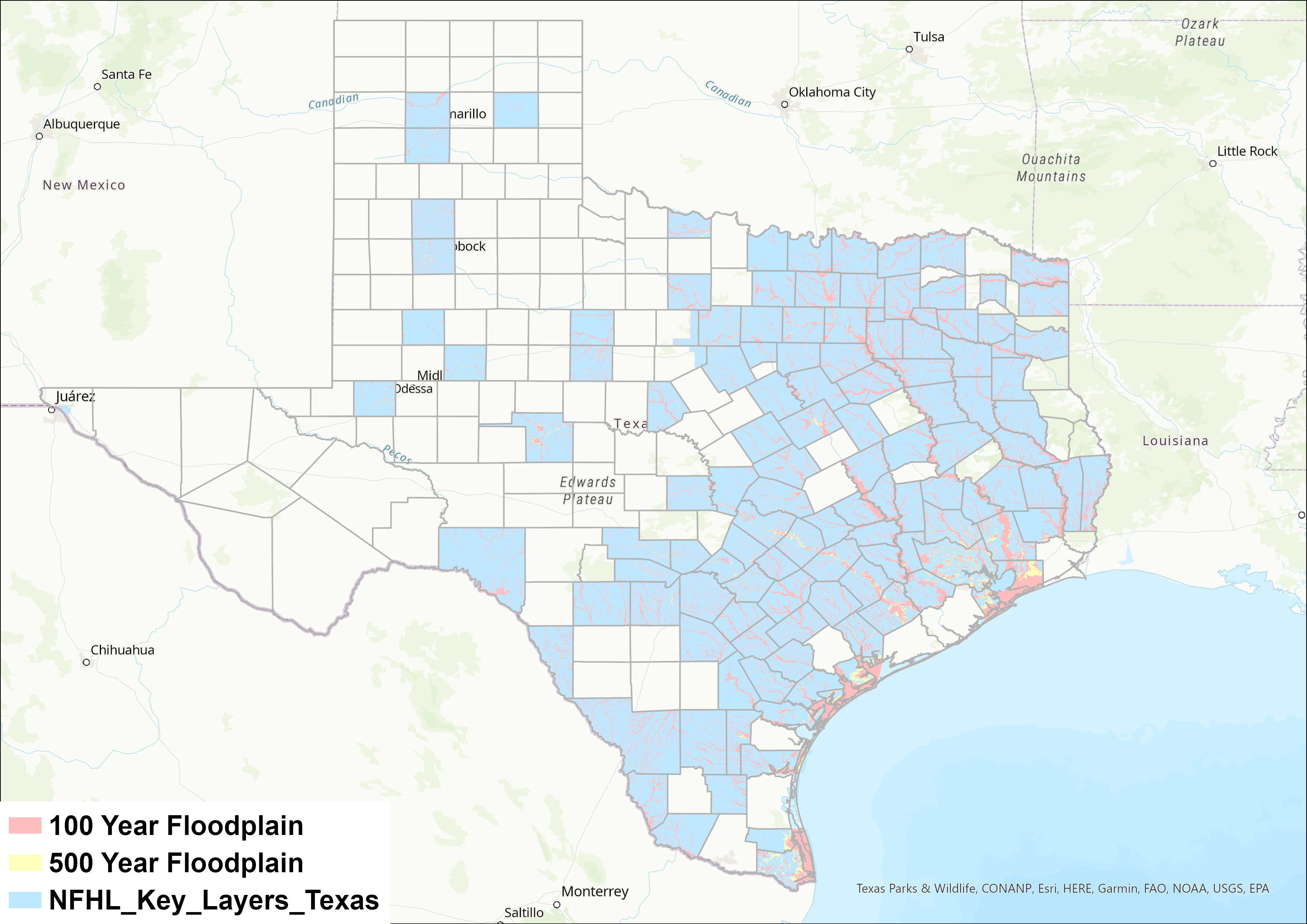}
    \caption{Floodplain data availability in Texas. The major urban areas in Texas, such as Houston, Dallas/Fort Worth, and Austin, have available floodplain data, whereas floodplain data is missing in some of the northwestern counties in Texas.}
	\label{fig:fig7}
\end{figure}

The floodplain data used in this case study was obtained from the NFHL provided by FEMA on Data.gov. However, floodplain data is available for only some districts, including major urban areas in Texas, and missing in some northwestern counties in Texas (Figure \ref{fig:fig7}). FEMA flood zones are geographic areas that FEMA has defined according to varying levels of flood risk. Each zone reflects the severity or type of flooding in the area based on the definition of flood zones, which distinguish the 100-year floodplain, 500-year floodplain, and others. Each of the 100-year and 500-year floodplains consists of several subcategories. The 100-year floodplain means there is a 1\% annual flooding probability and is identified as high risk, while the 500-year floodplain refers to a 0.2\% annual flooding probability and moderate risk. Regions located outside the 500-year floodplain will be identified as minimal risk or unknown risk.

\section{Proximity to Essential Facilities}
\label{sec:Proximity to Essential Facilities}
The previous two metrics evaluate the criticality of road segments based on network connectivity and vulnerability to flooding. Essential facilities, however, also play an important role during and in the aftermath of disasters; thus, road segments that are close to those essential facilities become critical because of the need to transport resources and to provide public services. This section evaluates the criticality of road segments based on the capability to provide access to critical facilities.

\subsection{Metric Definition and Calculation}
\begin{figure}
	\centering
    \includegraphics[width=0.95\linewidth]{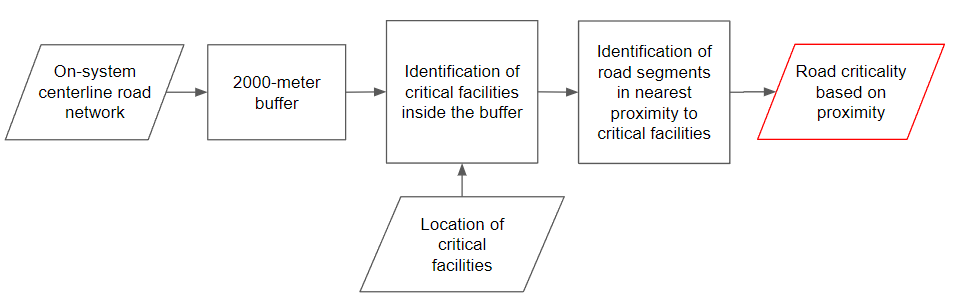}
    \caption{Steps for evaluating the criticality of road segments based on proximity to critical facilities.}
	\label{fig:fig8}
\end{figure}

To evaluate the criticality of road segments in road networks, we implemented the steps shown in Figure \ref{fig:fig8}. The idea is to evaluate the number of critical facilities to which a road segment provides direct access. We took the simplified road network as described in the previous section, calculated a 2,000-meter buffer around it, and identified the number of critical facilities that lie in this buffer. The purpose of this step is (1) to arrive at an estimated number of critical facilities distributed close to road networks and (2) to shortlist facilities to be used for computing the criticality metric. A buffer value of 2,000 meters was selected because most of the critical facilities lie within this buffer range. Moreover, we assumed that if a facility is within a 2,000-meter buffer range of a road network, it can be considered spatially co-located (co-location interdependency). This buffer value was selected based on our case study area; however, the buffer value may differ based on the density of the critical facilities in an area and the density of road infrastructure. After extracting the information of all the critical facilities that lie close to road networks, we computed the criticality metric by counting the number of facilities that lie in the nearest proximity (2,000-meter buffer range) of each road segment to evaluate the importance of each road segment. These road segments are then assigned a criticality score from level 1 to level 5, where level 1 is the most critical level for road segments and level 5 is the least critical level. In this study, if a road segment had more than ten critical facilities in its nearest proximity, it was assigned a score of level 1; if it had five to ten facilities, it was assigned to level 2; if it had two to five facilities, it was assigned to level 3; if it had only one facility, it was assigned to level 4; and if the road segment did not have any facility in its nearest proximity, it was assigned a score of level 5. The criticality levels of 1 and 2 mean that a road segment is critical and needs to be maintained for ensuring capable access to critical infrastructure in the vicinity, whereas levels 4 and 5 imply that there are none or less significant numbers of critical infrastructure in the nearest proximity to the road segment.

\subsection{Data Collection and Processing for the Case Study}
\begin{figure}
	\centering
    \includegraphics[width=0.95\linewidth]{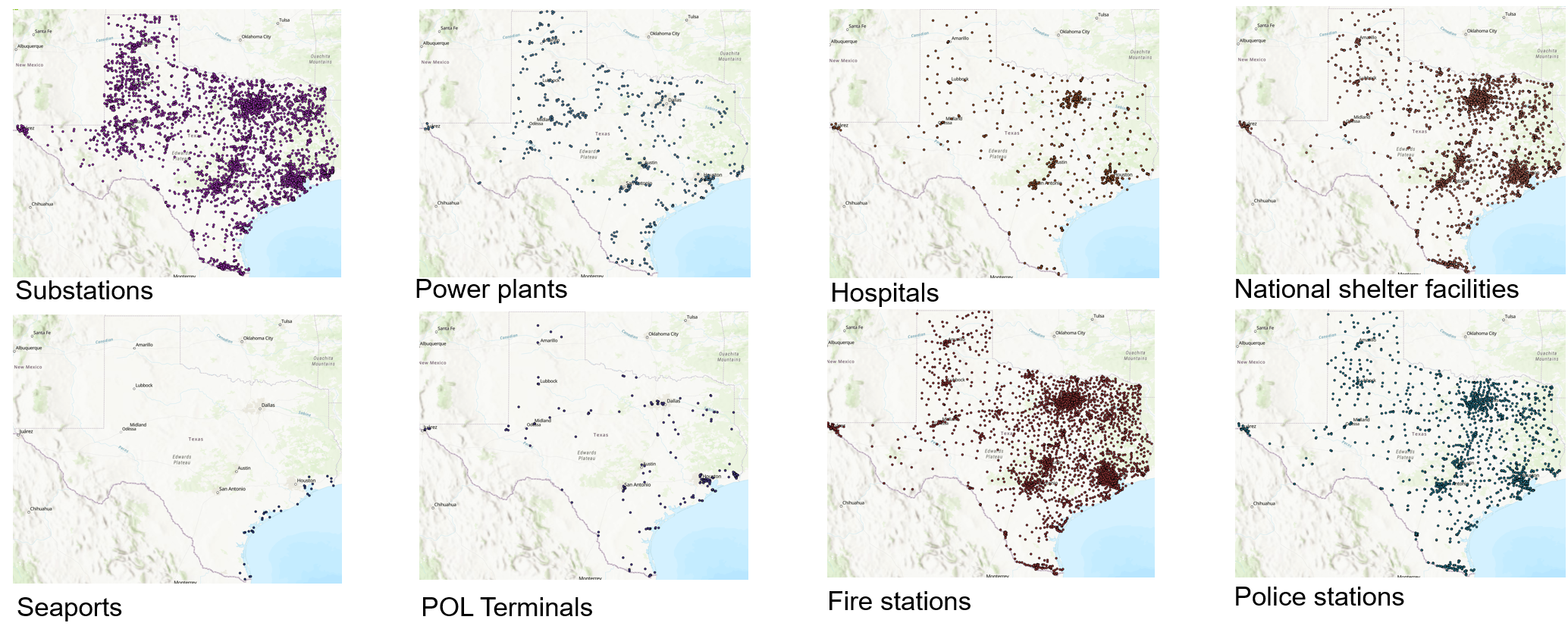}
    \caption{Spatial distribution of the eight critical facilities in the State of Texas.}
	\label{fig:fig9}
\end{figure}

In the case study, we collected location-based data of eight types of critical facilities, which are electrical substations, power plants, hospitals, national shelter facilities, seaports, fire stations, police stations, as well as petroleum, oil and lubricant (POL) terminals, from HIFLD and ArcGIS open data portal. The data consist of the coordinates of the respective critical facilities. It is important to maintain functionality of each of these facilities, including electricity, safety, health, and disaster relief. Access to critical facilities, especially during disruptions caused by disasters or other factors, is paramount for resiliency. Figure \ref{fig:fig9} shows the spatial distribution of the critical facilities in the state of Texas. Aggregating all the facilities together, we collected location information for around 15,000 critical facilities.

\section{Cascading Impact of Critical Infrastructure Networks}
\label{sec:Cascading Impact of Critical Infrastructure Networks}
\begin{figure}
	\centering
    \includegraphics[width=0.65\linewidth]{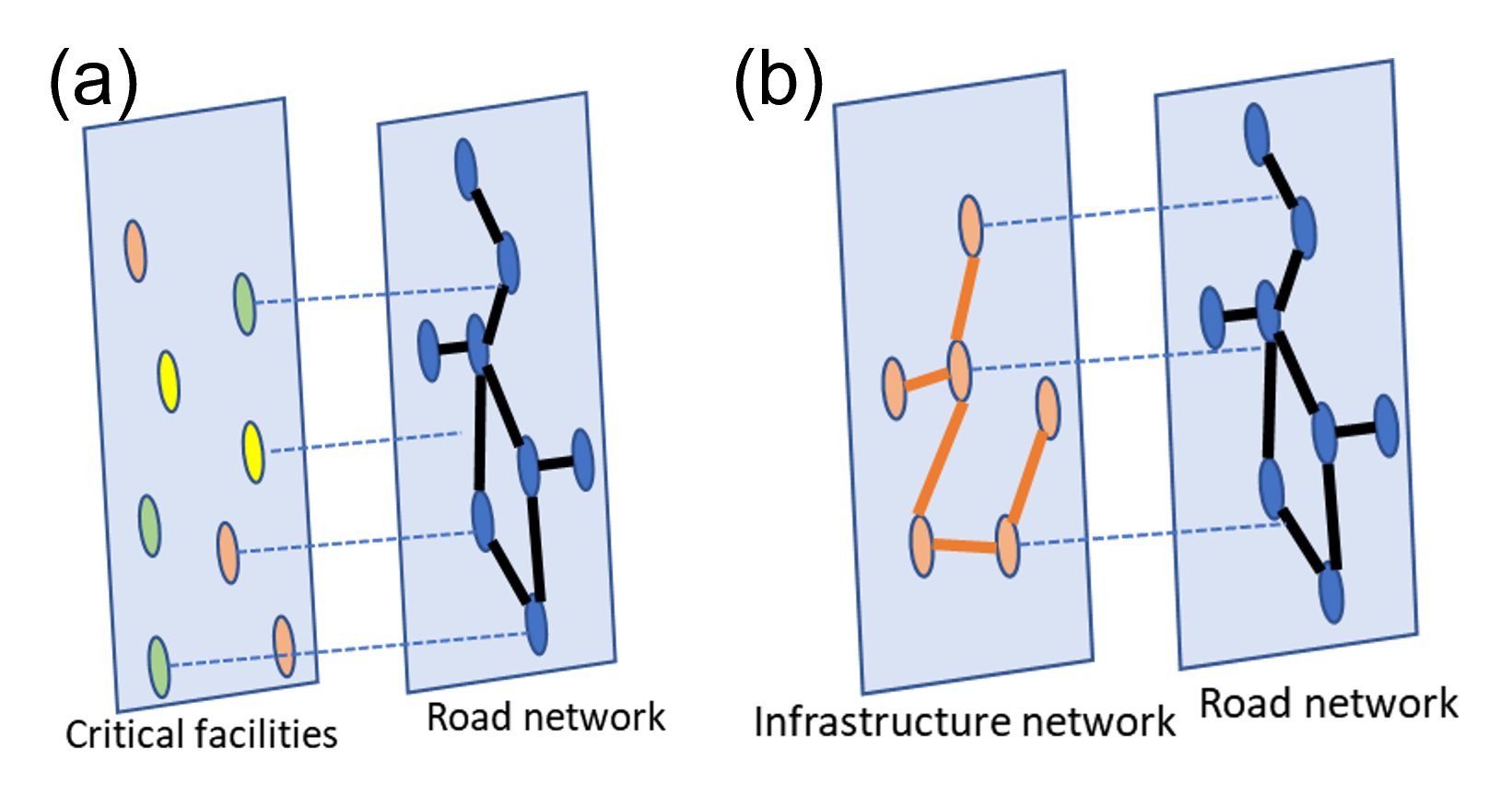}
    \caption{Schematic of assessing criticality of road segments based on (a)critical facilities and (b) critical facility networks.}
	\label{fig:fig10}
\end{figure}

The criticality metric proposed in the previous section, proximity to essential facilities, evaluates the criticality of road segments based on the locations of critical facilities that can be demonstrated as points on maps. Some critical infrastructures, however, such as the electrical power grid, railroads, and oil pipelines, are not conducive to be captured by the metric of proximity to essential facilities because they are spatially distributed as networks instead of single location attributes. Failure of a link within an infrastructure network can lead to severe cascading impacts on the entire infrastructure network \citep{fekete_critical_2019,lee_community-scale_2021,serre_assessing_2018}. For this reason, we calculated the criticality of road segments according to their closeness to critical infrastructure networks to understand the criticality related to cascading impacts. Figure \ref{fig:fig10}a shows the approach used in the previous metric where the criticality score is based on proximity to essential facilities, whereas Figure \ref{fig:fig10}b demonstrates how we calculated the criticality score based on cascading impact of critical infrastructure networks.

\subsection{Metric Definition and Calculation}
\begin{figure}
	\centering
    \includegraphics[width=0.95\linewidth]{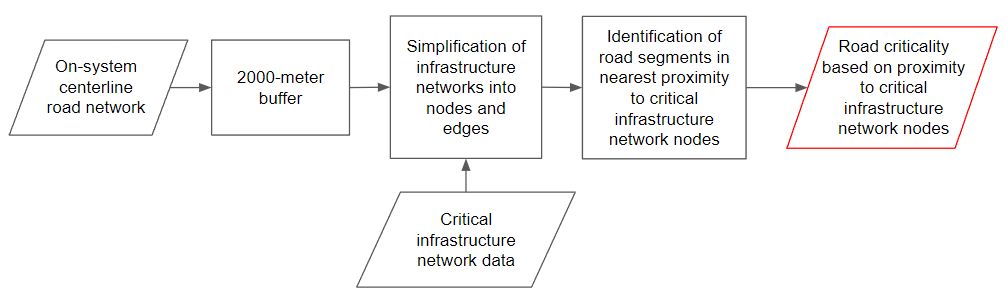}
    \caption{Steps for evaluating the criticality of road segments based on proximity to infrastructure networks.}
	\label{fig:fig11}
\end{figure}

To compute the criticality scores based on cascading impacts of critical infrastructure networks, we adopted an approach similar that described in Section \ref{sec:Proximity to Essential Facilities} in which we used the nodes of infrastructure networks as a proxy to determine the interdependency of infrastructure networks with road networks. Figure \ref{fig:fig11} shows the analysis steps used to compute this metric. Based on the criticality measure (i.e., counts of critical facility nodes), we assigned a criticality score from level 1 to level 5, where level 1 is the most critical level for road segments and level 5 is the least critical level. In this study, if a road segment had more than ten critical infrastructure network nodes in its nearest proximity, it was assigned a score of level 1; if it had five to ten infrastructure network nodes, it was assigned to level 2; if it had two to five infrastructure network nodes, it was assigned to level 3; if it had only one infrastructure network node, it was assigned to level 4; and if the road segment did not have any infrastructure network node in its nearest proximity, it was assigned a score of level 5.

\subsection{Data Collection and Processing for the Case Study}
\begin{figure}
	\centering
    \includegraphics[width=0.95\linewidth]{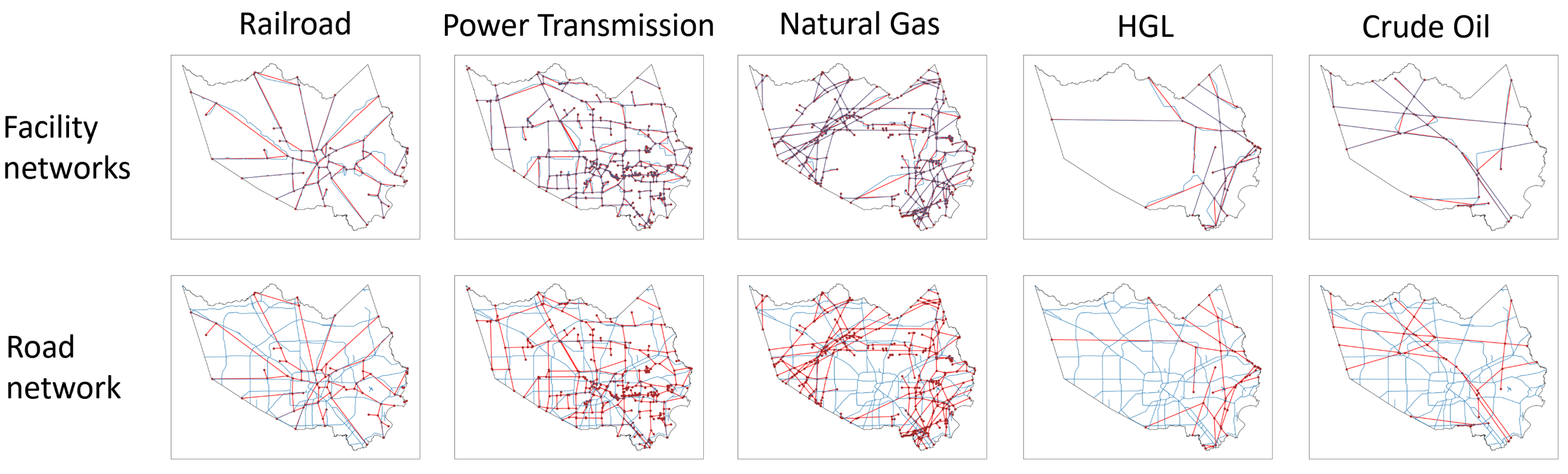}
    \caption{Example of the facility networks and road network in Harris County. Row 1: Infrastructure networks that were simplified to node-edge networks. Row 2: Mapping of simplified infrastructure networks with road networks.}
	\label{fig:fig12}
\end{figure}

We collected critical infrastructure network data for five types of critical infrastructure networks, which are railroad networks, transmission lines, gas lines, hydrocarbon gas lines (HGL), and oil pipelines. These datasets were acquired from sources such as HIFLD and ArcGIS open data portal. We simplified infrastructure networks into a node-and-link network model using two steps: (1) we created a buffer between 100 and 500 meters for each infrastructure network and merged overlapping features. This was done to ensure that multiple parallel lines in the network could be merged. (2) Centerlines for the buffered networks were computed, merging some parallel lines into one and simplifying some intersections where multiple nodes are combined into one. Figure \ref{fig:fig12} uses Harris County for illustration. Figure \ref{fig:fig12} (row 1) shows the approximation of converting infrastructure networks into nodes and edges. Figure \ref{fig:fig12} (row 2) shows superimposed simplified infrastructure networks with the road network in Harris County.

\section{Integrated Criticality Metric}
\label{sec:Integrated Criticality Metric}
After calculating the four criticality metrics we developed, the integrated criticality metrics that were calculated to help decision-makers understand the overall criticality of every road segment within the road network. While every criticality metric provides unique insight, a single integrated metric that comprises all information from different dimensions can be informative as well. In this section, we introduce two approaches to calculate metrics for the overall criticality scores and present the results of the integrated metrics. In addition, we compare the results of the two integrated criticality metrics and discuss the findings based on the comparison.

\subsection{Metric Development}
To understand the overall criticality of each road segments in networks, we calculated two integrated criticality metrics, which are based on evenly weighted sum (EWS) ranking and root squared Sum (RSS) ranking. The equations for EWS (Equation (\ref{eq:eq1})) and RSS (Equation (\ref{eq:eq2})) ranking calculations are: 

\begin{equation}
	EWS_{RS}= \sum {Level_{RS,CA}}
	\label{eq:eq1}
\end{equation}

\begin{equation}
	RSS_{RS}= \sqrt{\sum {{Level_{RS,CA}}^2}}
	\label{eq:eq2}
\end{equation}

where, $EWS_{RS}$ and $RSS_{RS}$ are the ranking values for each road segment $RS$ and $Level_{RS,CA}$ is the criticality levels of each road segment and all criticality aspects $CA$ examined. For instance, if a road segment has a criticality level of 3 for connectivity, level 2 for vulnerability to flooding, level 4 for proximity to essential facilities, and level 4 for cascading impact, the EWS ranking value of this road segment will be 13, whereas the RSS ranking value will be about 6.71. After ranking values for all road segments, these values are reclassified into five levels. We assigned an integrated criticality level to each road segment. The classification method divides the range of all ranking values (i.e., maximum subtracts minimum of ranking values of all road segments) into five equal intervals, representing five classes. The level 1 category, comprising the most critical segments in the network, is the highest 20\% of the ranking values. The class of level 5 category comprising the least critical segments in the network, is the lowest 20\% of the ranking values.

\section{Case Study for Texas Road Network}
\label{sec:Case Study for Texas Road Network}
This section presents the results for the four individual and two integrated metrics used to assess the criticality of road segments in the state of Texas. 

\subsection{Connectivity of Road Segments}
\begin{figure}
	\centering
    \includegraphics[width=0.95\linewidth]{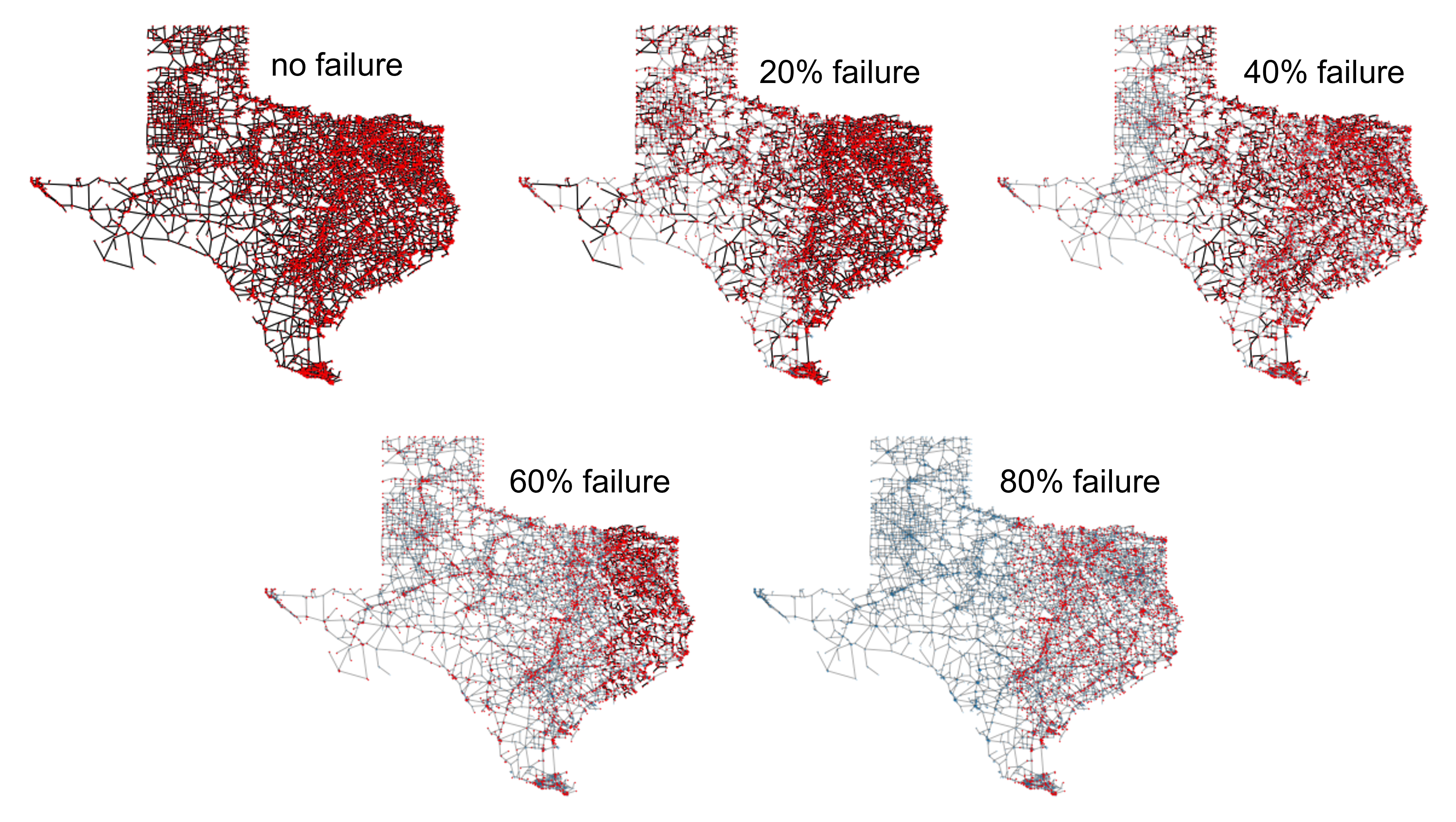}
    \caption{Remaining links after removing 20\% of nodes with the highest degrees and their corresponding failure links at each stage in Texas.}
	\label{fig:fig13}
\end{figure}

\begin{figure}
	\centering
    \includegraphics[width=0.5\linewidth]{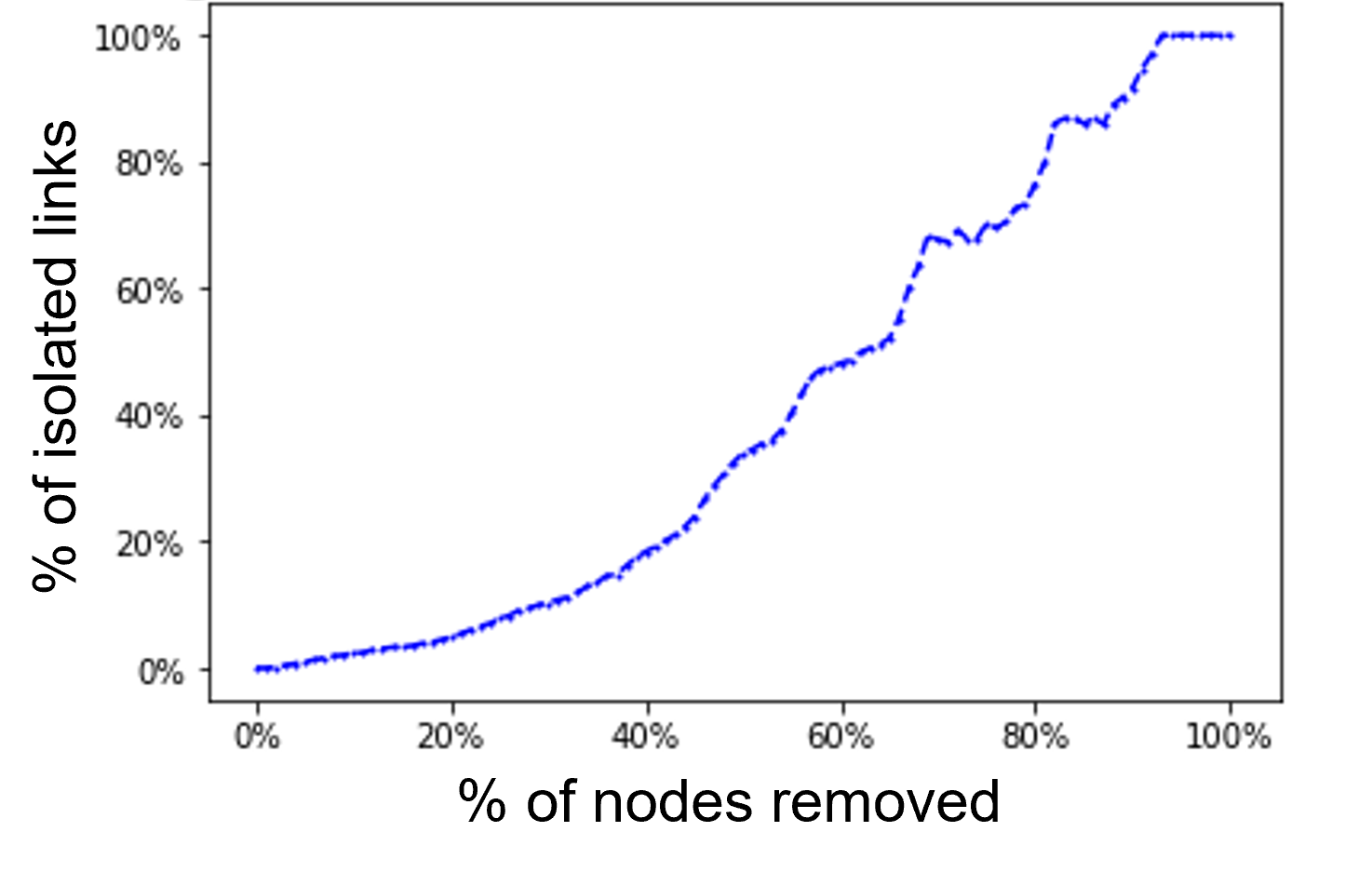}
    \caption{Percentage of links being isolated during the removal process. There is no abrupt change in the percentage of isolated links with the percentage change in nodes removed.}
	\label{fig:fig14}
\end{figure}

This metric reveals which road segments are critical in terms of connecting different parts of the network. The more critical a road segment is, the higher impact it will cause on overall connectivity when it is damaged; thus, this metric is a key indicator for informing road infrastructure planning. Figure \ref{fig:fig13} shows the remaining links after removing 20\% of nodes with the highest degrees at each stage. Figure \ref{fig:fig14} shows the percentage of links being isolated during the removal process. During the link removal process that simulates the failure of road segments, critical transitions (sudden increases in the percentage of links becoming isolated) are not observed, which indicates that no particular segment has substantial impacts to the network when failing. From observing the visual results in Texas, most of the road segments that belong to higher criticality levels (levels 1 and 2) are located in the outskirts of major urban areas or in coastal areas. These road segments may be part of the busy commuting routes or the weakly connected parts of the network. The criticality level for this metric depends solely on the topology of the road network. Figure \ref{fig:figa1} shows the criticality score of the road segments in Texas based on the connectivity metric. Most of the highly critical links, links with levels 1 and 2 criticalities, appear on the outskirts of major cities and in the coastal regions where fewer alternative substitute routes exist in case a link is removed.

\subsection{Flooding Vulnerability}
As aforementioned, flooding is the most prominent hazard in the state of Texas. This metric shows road segments that are more vulnerable to flooding-related hazards. Even though we made our best efforts to have floodplain data for all counties in Texas, floodplain data in some counties are not available, leaving the criticality of some road segments unclassified. In the case study, the road segments located in the 100- and 500-year floodplain, which are identified as high- and moderate-risk road segments by the definition of FEMA, are assigned to levels 1 and 2, and which form only about 20\% of the total number of road segments. The rest of the road segments are unlikely to be affected by flooding and are relatively less vulnerable. In the Appendix \ref{sec:Visualization Maps for the Case Study Area}, Figure \ref{fig:figa2} shows the criticality of road segments based on this metric. The road segments that are identified as levels 1 and 2 may need additional attention to enhance their resiliency and to plan for alternative routes while affected by the hazards. 

\subsection{Proximity to Essential Facilities}
This criticality metric captures the importance of road segments that can provide access to critical and essential facilities. We find that the critical road segments identified by this metric are more concentrated in the centers of the urban areas. In particular, the critical road segments are found in the high-density metropolitan areas of Houston, San Antonio, Austin, and Dallas/Fort Worth, which are the four major urban areas in Texas. Since the concentration of critical facilities is relative denser in urban and populated areas with respect to the road density, there are more critical road segments in terms of proximity to essential facilities present in urban areas. Figure \ref{fig:figa3} illustrates important road segments (levels 1 and 2) in Texas. The road segments having higher criticality in terms of proximity to essential facilities are located mainly in the major urban areas in Texas. Additionally, most level 3 road segments that have moderate criticality are located at major road junctions. On the other hand, rural areas consist mostly of low criticality road segments (levels 4 and 5) due to a lower concentration of critical facilities.

\subsection{Cascading Impact on Critical Infrastructure Networks}
The metric of cascading impact on critical infrastructure networks, evaluates the criticality of infrastructure networks based on their colocation interdependency with the road network. That is, the more infrastructure network nodes are in close proximity to the road segment, the higher the criticality of cascading impact on critical infrastructure networks is for a road segment. Here, unlike the metric of proximity to critical facilities, which demonstrates that the critical road segments are more concentrated in the urban areas, we found that most of the critical road segments in terms of cascading impact on critical infrastructure networks are near the coastline and western Texas instead of concentrated around urban areas. Moderately critical road segments are mostly present in urban areas. Figure \ref{fig:figa4} shows the critical road segments in red and orange (levels 1 and 2) based on their interdependency to critical infrastructure networks. 

\subsection{Integrated Criticality Metrics}
\begin{figure}
	\centering
    \includegraphics[width=0.6\linewidth]{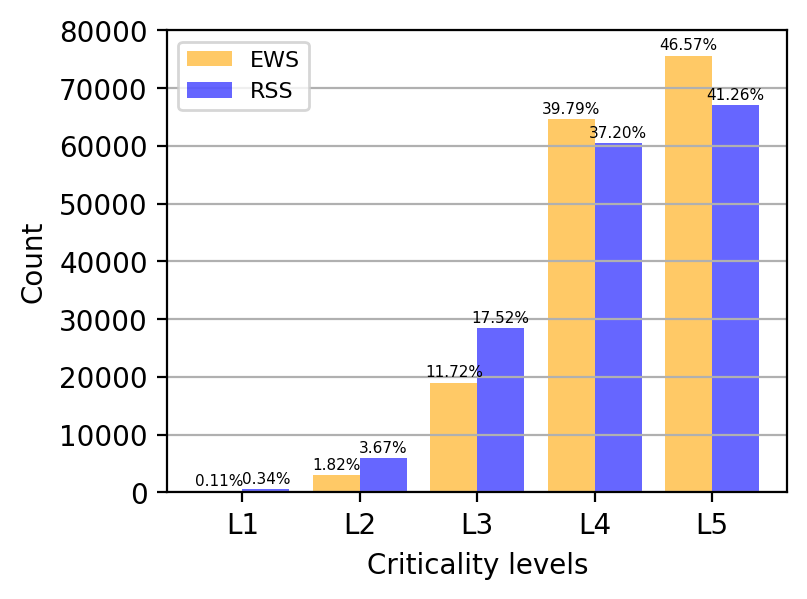}
    \caption{Distribution of different criticality levels of the RSS and EWS methods for the state of Texas. }
	\label{fig:fig15}
\end{figure}

\begin{figure}
	\centering
    \includegraphics[width=0.9\linewidth]{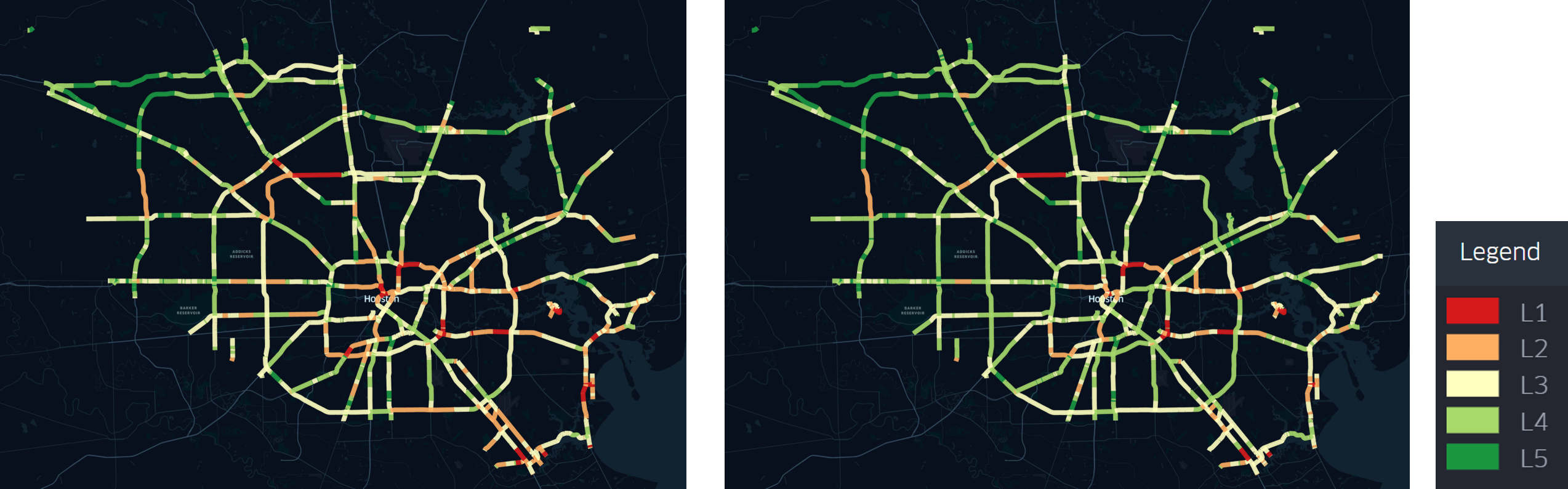}
    \caption{Results of EWS ranking (left) and RSS ranking (right) in Harris County.}
	\label{fig:fig16}
\end{figure}

The results of implementing the two integrated criticality metrics in Texas, as an example, are demonstrated in Figures \ref{fig:figa5} and \ref{fig:figa6}. More than 75\% of the road segments are classified as levels 4 and 5 for both integrated criticality metrics. As shown in Figure \ref{fig:fig15}, the EWS method tends to have more levels 1, 2, and 3 criticality road segments, which is more conservative than the RSS method. By implementing the integrated criticality metrics, the comparison of criticality levels can be conducted at any scale and location in Texas (e.g., metropolitan area and county level) to understand the overall criticality of each road segment. Figure \ref{fig:fig16} demonstrates an example in Harris County that the level 1 and 2 road segments are generally located close to intersections of road segments. In addition, the road segments at the northwestern corner of Harris County, where the city of Houston is located, have relatively minor criticality compared to the road segments in the other area of Harris County.

\begin{figure}
	\centering
    \includegraphics[width=0.8\linewidth]{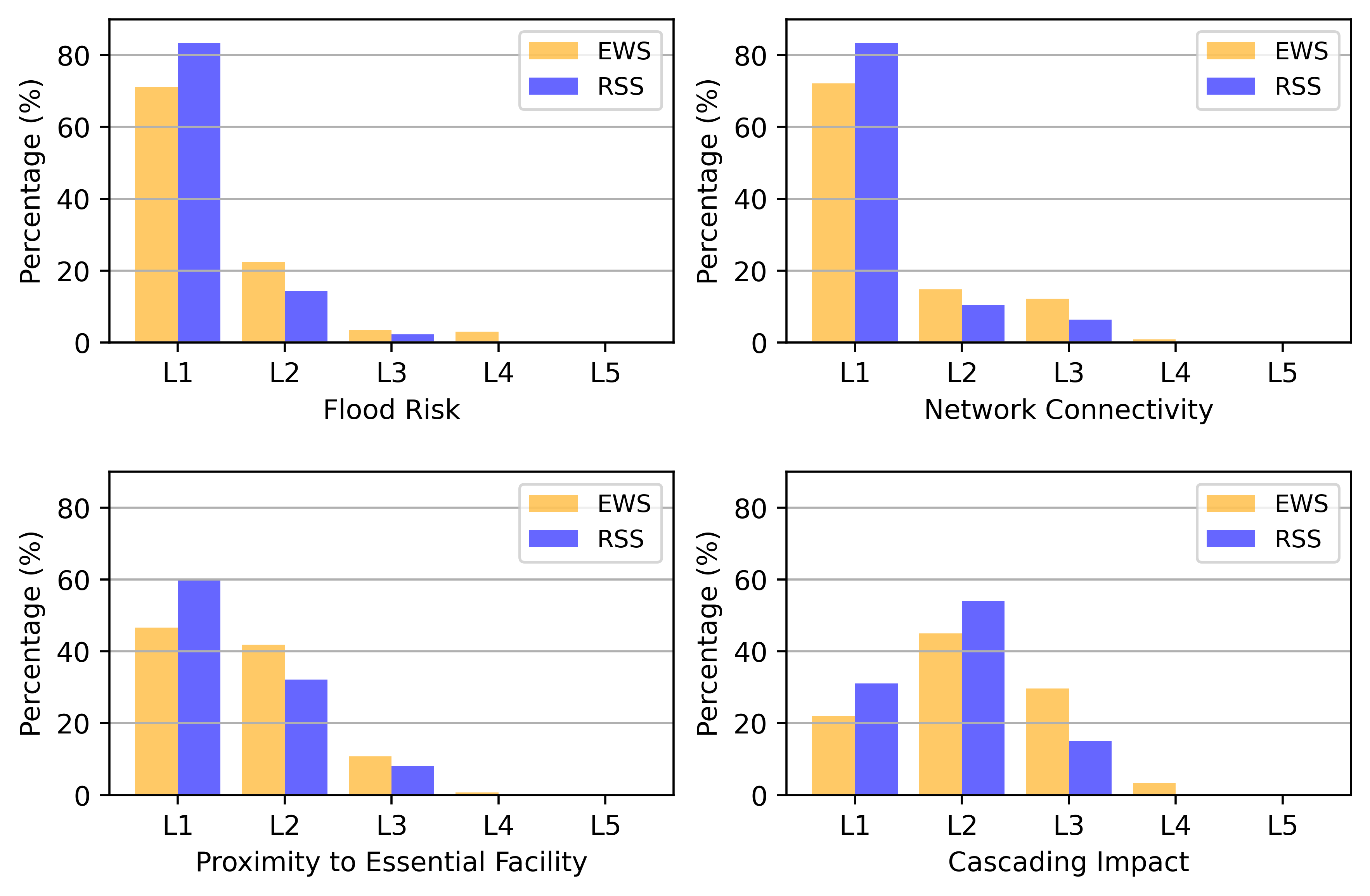}
    \caption{Distribution of the influence of individual metrics for the road segments with EWS and RSS integrated criticality of level 1.}
	\label{fig:fig17}
\end{figure}

Based on the integrated criticality metrics, we can specify the overall criticality of different road segments and prioritize resilience investments accordingly. Thus, it is essential to know the most common criticality concern among the four metrics if a road segment is classified as level 1 (i.e., the most critical level) for the overall criticality level. Figure \ref{fig:fig17} shows the distribution of the four individual metrics for the EWS and RSS integrated criticality metrics in level 1. More than 70\% of the road segments in the Texas road networks assigned an EWS integrated criticality of level 1 are critical (i.e., level 1 criticality) in terms of flood exposure and network connectivity. Similarly, more than 80\% of the road segments with RSS integrated criticality of level 1 are also critical in flood exposure and network connectivity. On the other hand, the criticality in terms of cascading impacts appears to be relatively less influential for the road segments with the EWS and RSS integrated criticality metrics of level 1, of which the criticality related to cascading impacts are mostly at level 2. With these results, agencies and organizations can prioritize maintenance plans proactively and ensure the required functionality and serviceability of road networks to mitigate the impact of disasters and recover from impact rapidly.

\section{Concluding Remarks}
\label{sec:Concluding Remarks}
The objective of the study is to propose a framework with quantitative measures incorporating resilience assessment into transportation planning and project development processes in practice. While the literature discusses several methods and measures for resilience assessment of transportation infrastructure, just a small fraction of the existing work has been adopted in practice to inform transportation planning and project development. This limitation is mainly due to two reasons: (1) transportation planners and decision makers do not know which to use among several methods and measures proposed in the literature; (2) the methods and measures for vulnerability and resilience assessment should be obtainable using the available data. To address this gap, in this study, we proposed and tested a framework for system-level assessment of transportation infrastructure networks based on four measures: (1) loss of connectivity of road segments in road networks, (2) vulnerability of road segments to extreme events, such as flooding and hurricanes, (3) disrupted access to critical facilities, and (4) cascading impacts of critical infrastructure networks. In addition, two integrated metrics were proposed to combine all metrics into one measure that can show the overall performance of road segments in terms of resilience. To demonstrate the use of the framework, we implemented individual and integrated metrics in the state of Texas. We collected datasets related to the topology of the Texas road network, hazard characteristics, and the location of critical facilities (e.g., power substations, oil/gas lines, and hospitals) from open sources and calculated the values of each metric for all road network links in Texas. The results showed that the framework can effectively provide useful insights with quantitative measures. In addition, the visualization of the results can make the interpretation of resilience assessment on road networks more straightforward and comprehensive. Also, the case study demonstrates the ability of the framework to be conducted at any scale and location to reveal the resilience and criticality of each road segment on networks.

The framework proposed in this paper can help transportation agencies incorporate resilience considerations into planning and project development proactively and to understand the resilience performance of the current road networks effectively. For example, road segments with high criticality and vulnerability levels could be prioritized for adaptation and hazard mitigation retrofit. Roads with high criticality and vulnerability to flooding impacts could be effectively identified and appropriate projects (such as drainage improvement or road elevation increase) could be implemented to reduce the vulnerability of those road segments. The demonstration of the results of this study to planners and decision makers in Texas has confirmed that the produced measures are deemed useful by practitioners. In fact, at the time of writing this manuscript, a state-wide transportation resilience planning effort is underway, and the results of this study are a major component in informing the planning process. Since the data used in this study is available to other states and countries, the framework presented in this study can be adopted by other transportation agencies across the globe for regional transportation resilience assessments. 

\section{Data availability}
All datasets used in this study are publicly available. The road network data can be acquired from the TxDOT data repository. Flood impact data is extracted from the National Flood Plain Layer on the FEMA website. Data on critical facilities and critical infrastructure networks can be accessed from publicly available Homeland Infrastructure Foundation Level Data (HIFLD) and ArcGIS open data portals.

\section{Code availability}
The code that supports the findings of this study is available from the corresponding author upon request.

\section{Acknowledgements}
This research is part of the study supported by the Texas Department of Transportation (TxDOT). The contents of this paper reflect the views of the authors, who are responsible for the facts and the accuracy of the data presented herein. The contents do not necessarily reflect the official view or policies of TxDOT. Any opinions, findings, and conclusions or recommendations expressed in this paper are those of the authors and do not necessarily reflect the views of the organizations or the individuals listed here.
\bibliography{ref}  

\begin{thebibliography}{}

\bibitem [\protect \citeauthoryear {%
Bruneau%
\ \protect \BOthers {.}}{%
Bruneau%
\ \protect \BOthers {.}}{%
{\protect \APACyear {2003}}%
}]{%
bruneau_framework_2003}
\APACinsertmetastar {%
bruneau_framework_2003}%
\begin{APACrefauthors}%
Bruneau, M.%
, Chang, S\BPBI E.%
, Eguchi, R\BPBI T.%
, Lee, G\BPBI C.%
, O'Rourke, T\BPBI D.%
, Reinhorn, A\BPBI M.%
\BDBL {}von Winterfeldt, D.%
\end{APACrefauthors}%
\unskip\
\newblock
\APACrefYearMonthDay{2003}{{\APACmonth{11}}}{}.
\newblock
{\BBOQ}\APACrefatitle {A {Framework} to {Quantitatively} {Assess} and {Enhance}
  the {Seismic} {Resilience} of {Communities}} {A {Framework} to
  {Quantitatively} {Assess} and {Enhance} the {Seismic} {Resilience} of
  {Communities}}.{\BBCQ}
\newblock
\APACjournalVolNumPages{Earthquake Spectra}{19}{4}{733--752}.
\newblock
\begin{APACrefURL} [{2022-03-25}]\url{https://doi.org/10.1193/1.1623497}
  \end{APACrefURL}
\newblock
\APACrefnote{Publisher: SAGE Publications Ltd STM}
\newblock
\begin{APACrefDOI} \doi{10.1193/1.1623497} \end{APACrefDOI}
\PrintBackRefs{\CurrentBib}

\bibitem [\protect \citeauthoryear {%
Da~Silva%
\ \BBA {} Morera%
}{%
Da~Silva%
\ \BBA {} Morera%
}{%
{\protect \APACyear {2014}}%
}]{%
da_silva_city_2014}
\APACinsertmetastar {%
da_silva_city_2014}%
\begin{APACrefauthors}%
Da~Silva, J.%
\BCBT {}\ \BBA {} Morera, B.%
\end{APACrefauthors}%
\unskip\
\newblock
\APACrefYearMonthDay{2014}{}{}.
\newblock
{\BBOQ}\APACrefatitle {City resilience framework} {City resilience
  framework}.{\BBCQ}
\newblock
\APACjournalVolNumPages{London: Arup}{}{}{}.
\PrintBackRefs{\CurrentBib}

\bibitem [\protect \citeauthoryear {%
Dave%
, Subramanian%
\BCBL {}\ \BBA {} Bhatia%
}{%
Dave%
\ \protect \BOthers {.}}{%
{\protect \APACyear {2021}}%
}]{%
dave_extreme_2021}
\APACinsertmetastar {%
dave_extreme_2021}%
\begin{APACrefauthors}%
Dave, R.%
, Subramanian, S\BPBI S.%
\BCBL {}\ \BBA {} Bhatia, U.%
\end{APACrefauthors}%
\unskip\
\newblock
\APACrefYearMonthDay{2021}{{\APACmonth{10}}}{}.
\newblock
{\BBOQ}\APACrefatitle {Extreme precipitation induced concurrent events trigger
  prolonged disruptions in regional road networks} {Extreme precipitation
  induced concurrent events trigger prolonged disruptions in regional road
  networks}.{\BBCQ}
\newblock
\APACjournalVolNumPages{Environmental Research Letters}{16}{10}{104050}.
\newblock
\begin{APACrefURL} [{2022-04-04}]\url{https://doi.org/10.1088/1748-9326/ac2d67}
  \end{APACrefURL}
\newblock
\APACrefnote{Publisher: IOP Publishing}
\newblock
\begin{APACrefDOI} \doi{10.1088/1748-9326/ac2d67} \end{APACrefDOI}
\PrintBackRefs{\CurrentBib}

\bibitem [\protect \citeauthoryear {%
Davis%
, Mostafavi%
\BCBL {}\ \BBA {} Wang%
}{%
Davis%
\ \protect \BOthers {.}}{%
{\protect \APACyear {2018}}%
}]{%
davis_establishing_2018}
\APACinsertmetastar {%
davis_establishing_2018}%
\begin{APACrefauthors}%
Davis, C\BPBI A.%
, Mostafavi, A.%
\BCBL {}\ \BBA {} Wang, H.%
\end{APACrefauthors}%
\unskip\
\newblock
\APACrefYearMonthDay{2018}{{\APACmonth{11}}}{}.
\newblock
{\BBOQ}\APACrefatitle {Establishing {Characteristics} to {Operationalize}
  {Resilience} for {Lifeline} {Systems}} {Establishing {Characteristics} to
  {Operationalize} {Resilience} for {Lifeline} {Systems}}.{\BBCQ}
\newblock
\APACjournalVolNumPages{Natural Hazards Review}{19}{4}{04018014}.
\newblock
\begin{APACrefURL}
  [{2022-03-25}]\url{https://ascelibrary.org/doi/full/10.1061/%28ASCE%29NH.1527-6996.0000303}
  \end{APACrefURL}
\newblock
\APACrefnote{Publisher: American Society of Civil Engineers}
\newblock
\begin{APACrefDOI} \doi{10.1061/(ASCE)NH.1527-6996.0000303} \end{APACrefDOI}
\PrintBackRefs{\CurrentBib}

\bibitem [\protect \citeauthoryear {%
Dix%
, Zgoda%
, Vargo%
, Heitsch%
\BCBL {}\ \BBA {} Gestwick%
}{%
Dix%
\ \protect \BOthers {.}}{%
{\protect \APACyear {2018}}%
}]{%
dix_integrating_2018}
\APACinsertmetastar {%
dix_integrating_2018}%
\begin{APACrefauthors}%
Dix, B.%
, Zgoda, B.%
, Vargo, A.%
, Heitsch, S.%
\BCBL {}\ \BBA {} Gestwick, T.%
\end{APACrefauthors}%
\unskip\
\newblock
\APACrefYearMonthDay{2018}{{\APACmonth{05}}}{}.
\newblock
\APACrefbtitle {Integrating {Resilience} into the {Transportation} {Planning}
  {Process}: {White} {Paper} on {Literature} {Review} {Findings}} {Integrating
  {Resilience} into the {Transportation} {Planning} {Process}: {White} {Paper}
  on {Literature} {Review} {Findings}}\ \APACbVolEdTR{}{\BTR{}}.
\newblock
\begin{APACrefURL} [{2022-03-25}]\url{https://trid.trb.org/view/1516130}
  \end{APACrefURL}
\newblock
\APACrefnote{Number: FHWA-HEP-18-050}
\PrintBackRefs{\CurrentBib}

\bibitem [\protect \citeauthoryear {%
Dong%
, Esmalian%
, Farahmand%
\BCBL {}\ \BBA {} Mostafavi%
}{%
Dong%
, Esmalian%
\BCBL {}\ \protect \BOthers {.}}{%
{\protect \APACyear {2020}}%
}]{%
dong_integrated_2020}
\APACinsertmetastar {%
dong_integrated_2020}%
\begin{APACrefauthors}%
Dong, S.%
, Esmalian, A.%
, Farahmand, H.%
\BCBL {}\ \BBA {} Mostafavi, A.%
\end{APACrefauthors}%
\unskip\
\newblock
\APACrefYearMonthDay{2020}{{\APACmonth{03}}}{}.
\newblock
{\BBOQ}\APACrefatitle {An integrated physical-social analysis of disrupted
  access to critical facilities and community service-loss tolerance in urban
  flooding} {An integrated physical-social analysis of disrupted access to
  critical facilities and community service-loss tolerance in urban
  flooding}.{\BBCQ}
\newblock
\APACjournalVolNumPages{Computers, Environment and Urban
  Systems}{80}{}{101443}.
\newblock
\begin{APACrefURL}
  [{2022-03-25}]\url{https://www.sciencedirect.com/science/article/pii/S019897151930434X}
  \end{APACrefURL}
\newblock
\begin{APACrefDOI} \doi{10.1016/j.compenvurbsys.2019.101443} \end{APACrefDOI}
\PrintBackRefs{\CurrentBib}

\bibitem [\protect \citeauthoryear {%
Dong%
, Mostafizi%
, Wang%
, Gao%
\BCBL {}\ \BBA {} Li%
}{%
Dong%
, Mostafizi%
\BCBL {}\ \protect \BOthers {.}}{%
{\protect \APACyear {2020}}%
}]{%
dong_measuring_2020}
\APACinsertmetastar {%
dong_measuring_2020}%
\begin{APACrefauthors}%
Dong, S.%
, Mostafizi, A.%
, Wang, H.%
, Gao, J.%
\BCBL {}\ \BBA {} Li, X.%
\end{APACrefauthors}%
\unskip\
\newblock
\APACrefYearMonthDay{2020}{{\APACmonth{06}}}{}.
\newblock
{\BBOQ}\APACrefatitle {Measuring the {Topological} {Robustness} of
  {Transportation} {Networks} to {Disaster}-{Induced} {Failures}: {A}
  {Percolation} {Approach}} {Measuring the {Topological} {Robustness} of
  {Transportation} {Networks} to {Disaster}-{Induced} {Failures}: {A}
  {Percolation} {Approach}}.{\BBCQ}
\newblock
\APACjournalVolNumPages{Journal of Infrastructure Systems}{26}{2}{04020009}.
\newblock
\begin{APACrefURL}
  [{2022-03-25}]\url{https://ascelibrary.org/doi/full/10.1061/%28ASCE%29IS.1943-555X.0000533}
  \end{APACrefURL}
\newblock
\APACrefnote{Publisher: American Society of Civil Engineers}
\newblock
\begin{APACrefDOI} \doi{10.1061/(ASCE)IS.1943-555X.0000533} \end{APACrefDOI}
\PrintBackRefs{\CurrentBib}

\bibitem [\protect \citeauthoryear {%
Dong%
, Wang%
, Mostafavi%
\BCBL {}\ \BBA {} Gao%
}{%
Dong%
\ \protect \BOthers {.}}{%
{\protect \APACyear {2019}}%
}]{%
dong_robust_2019}
\APACinsertmetastar {%
dong_robust_2019}%
\begin{APACrefauthors}%
Dong, S.%
, Wang, H.%
, Mostafavi, A.%
\BCBL {}\ \BBA {} Gao, J.%
\end{APACrefauthors}%
\unskip\
\newblock
\APACrefYearMonthDay{2019}{{\APACmonth{08}}}{}.
\newblock
{\BBOQ}\APACrefatitle {Robust component: a robustness measure that incorporates
  access to critical facilities under disruptions} {Robust component: a
  robustness measure that incorporates access to critical facilities under
  disruptions}.{\BBCQ}
\newblock
\APACjournalVolNumPages{Journal of The Royal Society
  Interface}{16}{157}{20190149}.
\newblock
\begin{APACrefURL}
  [{2022-03-25}]\url{https://royalsocietypublishing.org/doi/full/10.1098/rsif.2019.0149}
  \end{APACrefURL}
\newblock
\APACrefnote{Publisher: Royal Society}
\newblock
\begin{APACrefDOI} \doi{10.1098/rsif.2019.0149} \end{APACrefDOI}
\PrintBackRefs{\CurrentBib}

\bibitem [\protect \citeauthoryear {%
Esmalian%
\ \protect \BOthers {.}}{%
Esmalian%
\ \protect \BOthers {.}}{%
{\protect \APACyear {2022}}%
}]{%
esmalian_operationalizing_2022}
\APACinsertmetastar {%
esmalian_operationalizing_2022}%
\begin{APACrefauthors}%
Esmalian, A.%
, Yuan, F.%
, Rajput, A\BPBI A.%
, Farahmand, H.%
, Dong, S.%
, Li, Q.%
\BDBL {}Mostafavi, A.%
\end{APACrefauthors}%
\unskip\
\newblock
\APACrefYearMonthDay{2022}{{\APACmonth{03}}}{}.
\newblock
{\BBOQ}\APACrefatitle {Operationalizing resilience practices in transportation
  infrastructure planning and project development} {Operationalizing resilience
  practices in transportation infrastructure planning and project
  development}.{\BBCQ}
\newblock
\APACjournalVolNumPages{Transportation Research Part D: Transport and
  Environment}{104}{}{103214}.
\newblock
\begin{APACrefURL}
  [{2022-03-25}]\url{https://www.sciencedirect.com/science/article/pii/S136192092200044X}
  \end{APACrefURL}
\newblock
\begin{APACrefDOI} \doi{10.1016/j.trd.2022.103214} \end{APACrefDOI}
\PrintBackRefs{\CurrentBib}

\bibitem [\protect \citeauthoryear {%
Fekete%
}{%
Fekete%
}{%
{\protect \APACyear {2019}}%
}]{%
fekete_critical_2019}
\APACinsertmetastar {%
fekete_critical_2019}%
\begin{APACrefauthors}%
Fekete, A.%
\end{APACrefauthors}%
\unskip\
\newblock
\APACrefYearMonthDay{2019}{}{}.
\newblock
{\BBOQ}\APACrefatitle {Critical infrastructure and flood resilience:
  {Cascading} effects beyond water} {Critical infrastructure and flood
  resilience: {Cascading} effects beyond water}.{\BBCQ}
\newblock
\APACjournalVolNumPages{WIREs Water}{6}{5}{e1370}.
\newblock
\begin{APACrefURL}
  [{2022-03-25}]\url{https://onlinelibrary.wiley.com/doi/abs/10.1002/wat2.1370}
  \end{APACrefURL}
\newblock
\APACrefnote{\_eprint:
  https://onlinelibrary.wiley.com/doi/pdf/10.1002/wat2.1370}
\newblock
\begin{APACrefDOI} \doi{10.1002/wat2.1370} \end{APACrefDOI}
\PrintBackRefs{\CurrentBib}

\bibitem [\protect \citeauthoryear {%
Flannery%
, Pena%
\BCBL {}\ \BBA {} Manns%
}{%
Flannery%
\ \protect \BOthers {.}}{%
{\protect \APACyear {2018}}%
}]{%
flannery_resilience_2018}
\APACinsertmetastar {%
flannery_resilience_2018}%
\begin{APACrefauthors}%
Flannery, A.%
, Pena, M\BPBI A.%
\BCBL {}\ \BBA {} Manns, J.%
\end{APACrefauthors}%
\ (\BEDS).
\unskip\
\newblock
\APACrefYear{2018}.
\newblock
\APACrefbtitle {Resilience in {Transportation} {Planning}, {Engineering},
  {Management}, {Policy}, and {Administration}} {Resilience in {Transportation}
  {Planning}, {Engineering}, {Management}, {Policy}, and {Administration}}.
\newblock
\APACaddressPublisher{Washington, DC}{The National Academies Press}.
\newblock
\begin{APACrefURL}
  [{2022-03-25}]\url{https://www.nap.edu/catalog/25166/resilience-in-transportation-planning-engineering-management-policy-and-administration}
  \end{APACrefURL}
\newblock
\begin{APACrefDOI} \doi{10.17226/25166} \end{APACrefDOI}
\PrintBackRefs{\CurrentBib}

\bibitem [\protect \citeauthoryear {%
Frizzelle%
, Evenson%
, Rodriguez%
\BCBL {}\ \BBA {} Laraia%
}{%
Frizzelle%
\ \protect \BOthers {.}}{%
{\protect \APACyear {2009}}%
}]{%
frizzelle_importance_2009}
\APACinsertmetastar {%
frizzelle_importance_2009}%
\begin{APACrefauthors}%
Frizzelle, B\BPBI G.%
, Evenson, K\BPBI R.%
, Rodriguez, D\BPBI A.%
\BCBL {}\ \BBA {} Laraia, B\BPBI A.%
\end{APACrefauthors}%
\unskip\
\newblock
\APACrefYearMonthDay{2009}{{\APACmonth{05}}}{}.
\newblock
{\BBOQ}\APACrefatitle {The importance of accurate road data for spatial
  applications in public health: customizing a road network} {The importance of
  accurate road data for spatial applications in public health: customizing a
  road network}.{\BBCQ}
\newblock
\APACjournalVolNumPages{International Journal of Health Geographics}{8}{1}{24}.
\newblock
\begin{APACrefURL} [{2022-04-04}]\url{https://doi.org/10.1186/1476-072X-8-24}
  \end{APACrefURL}
\newblock
\begin{APACrefDOI} \doi{10.1186/1476-072X-8-24} \end{APACrefDOI}
\PrintBackRefs{\CurrentBib}

\bibitem [\protect \citeauthoryear {%
Gao%
, Barzel%
\BCBL {}\ \BBA {} Barabási%
}{%
Gao%
\ \protect \BOthers {.}}{%
{\protect \APACyear {2016}}%
}]{%
gao_universal_2016}
\APACinsertmetastar {%
gao_universal_2016}%
\begin{APACrefauthors}%
Gao, J.%
, Barzel, B.%
\BCBL {}\ \BBA {} Barabási, A\BHBI L.%
\end{APACrefauthors}%
\unskip\
\newblock
\APACrefYearMonthDay{2016}{{\APACmonth{02}}}{}.
\newblock
{\BBOQ}\APACrefatitle {Universal resilience patterns in complex networks}
  {Universal resilience patterns in complex networks}.{\BBCQ}
\newblock
\APACjournalVolNumPages{Nature}{530}{7590}{307--312}.
\newblock
\begin{APACrefURL}
  [{2022-03-25}]\url{https://www.nature.com/articles/nature16948}
  \end{APACrefURL}
\newblock
\APACrefnote{Number: 7590 Publisher: Nature Publishing Group}
\newblock
\begin{APACrefDOI} \doi{10.1038/nature16948} \end{APACrefDOI}
\PrintBackRefs{\CurrentBib}

\bibitem [\protect \citeauthoryear {%
Hsieh%
\ \BBA {} Feng%
}{%
Hsieh%
\ \BBA {} Feng%
}{%
{\protect \APACyear {2020}}%
}]{%
hsieh_highway_2020}
\APACinsertmetastar {%
hsieh_highway_2020}%
\begin{APACrefauthors}%
Hsieh, C\BHBI H.%
\BCBT {}\ \BBA {} Feng, C\BHBI M.%
\end{APACrefauthors}%
\unskip\
\newblock
\APACrefYearMonthDay{2020}{{\APACmonth{03}}}{}.
\newblock
{\BBOQ}\APACrefatitle {The highway resilience and vulnerability in {Taiwan}}
  {The highway resilience and vulnerability in {Taiwan}}.{\BBCQ}
\newblock
\APACjournalVolNumPages{Transport Policy}{87}{}{1--9}.
\newblock
\begin{APACrefURL}
  [{2022-03-25}]\url{https://www.sciencedirect.com/science/article/pii/S0967070X18306292}
  \end{APACrefURL}
\newblock
\begin{APACrefDOI} \doi{10.1016/j.tranpol.2018.08.010} \end{APACrefDOI}
\PrintBackRefs{\CurrentBib}

\bibitem [\protect \citeauthoryear {%
Jenelius%
\ \BBA {} Mattsson%
}{%
Jenelius%
\ \BBA {} Mattsson%
}{%
{\protect \APACyear {2015}}%
}]{%
jenelius_road_2015}
\APACinsertmetastar {%
jenelius_road_2015}%
\begin{APACrefauthors}%
Jenelius, E.%
\BCBT {}\ \BBA {} Mattsson, L\BHBI G.%
\end{APACrefauthors}%
\unskip\
\newblock
\APACrefYearMonthDay{2015}{{\APACmonth{01}}}{}.
\newblock
{\BBOQ}\APACrefatitle {Road network vulnerability analysis:
  {Conceptualization}, implementation and application} {Road network
  vulnerability analysis: {Conceptualization}, implementation and
  application}.{\BBCQ}
\newblock
\APACjournalVolNumPages{Computers, Environment and Urban
  Systems}{49}{}{136--147}.
\newblock
\begin{APACrefURL}
  [{2022-04-04}]\url{https://www.sciencedirect.com/science/article/pii/S0198971514000192}
  \end{APACrefURL}
\newblock
\begin{APACrefDOI} \doi{10.1016/j.compenvurbsys.2014.02.003} \end{APACrefDOI}
\PrintBackRefs{\CurrentBib}

\bibitem [\protect \citeauthoryear {%
Jenelius%
, Petersen%
\BCBL {}\ \BBA {} Mattsson%
}{%
Jenelius%
\ \protect \BOthers {.}}{%
{\protect \APACyear {2006}}%
}]{%
jenelius_importance_2006}
\APACinsertmetastar {%
jenelius_importance_2006}%
\begin{APACrefauthors}%
Jenelius, E.%
, Petersen, T.%
\BCBL {}\ \BBA {} Mattsson, L\BHBI G.%
\end{APACrefauthors}%
\unskip\
\newblock
\APACrefYearMonthDay{2006}{{\APACmonth{08}}}{}.
\newblock
{\BBOQ}\APACrefatitle {Importance and exposure in road network vulnerability
  analysis} {Importance and exposure in road network vulnerability
  analysis}.{\BBCQ}
\newblock
\APACjournalVolNumPages{Transportation Research Part A: Policy and
  Practice}{40}{7}{537--560}.
\newblock
\begin{APACrefURL}
  [{2022-04-04}]\url{https://www.sciencedirect.com/science/article/pii/S096585640500162X}
  \end{APACrefURL}
\newblock
\begin{APACrefDOI} \doi{10.1016/j.tra.2005.11.003} \end{APACrefDOI}
\PrintBackRefs{\CurrentBib}

\bibitem [\protect \citeauthoryear {%
Lee%
, Maron%
\BCBL {}\ \BBA {} Mostafavi%
}{%
Lee%
\ \protect \BOthers {.}}{%
{\protect \APACyear {2021}}%
}]{%
lee_community-scale_2021}
\APACinsertmetastar {%
lee_community-scale_2021}%
\begin{APACrefauthors}%
Lee, C\BHBI C.%
, Maron, M.%
\BCBL {}\ \BBA {} Mostafavi, A.%
\end{APACrefauthors}%
\unskip\
\newblock
\APACrefYearMonthDay{2021}{{\APACmonth{08}}}{}.
\newblock
{\BBOQ}\APACrefatitle {Community-scale {Big} {Data} {Reveals} {Disparate}
  {Impacts} of the {Texas} {Winter} {Storm} of 2021 and its {Managed} {Power}
  {Outage}} {Community-scale {Big} {Data} {Reveals} {Disparate} {Impacts} of
  the {Texas} {Winter} {Storm} of 2021 and its {Managed} {Power}
  {Outage}}.{\BBCQ}
\newblock
\APACjournalVolNumPages{arXiv:2108.06046 [physics]}{}{}{}.
\newblock
\begin{APACrefURL} [{2021-08-18}]\url{http://arxiv.org/abs/2108.06046}
  \end{APACrefURL}
\newblock
\APACrefnote{ZSCC: NoCitationData[s0] arXiv: 2108.06046}
\PrintBackRefs{\CurrentBib}

\bibitem [\protect \citeauthoryear {%
Martín%
, Ortega%
, Cuevas-Wizner%
, Ledda%
\BCBL {}\ \BBA {} De~Montis%
}{%
Martín%
\ \protect \BOthers {.}}{%
{\protect \APACyear {2021}}%
}]{%
martin_assessing_2021}
\APACinsertmetastar {%
martin_assessing_2021}%
\begin{APACrefauthors}%
Martín, B.%
, Ortega, E.%
, Cuevas-Wizner, R.%
, Ledda, A.%
\BCBL {}\ \BBA {} De~Montis, A.%
\end{APACrefauthors}%
\unskip\
\newblock
\APACrefYearMonthDay{2021}{{\APACmonth{06}}}{}.
\newblock
{\BBOQ}\APACrefatitle {Assessing road network resilience: {An} accessibility
  comparative analysis} {Assessing road network resilience: {An} accessibility
  comparative analysis}.{\BBCQ}
\newblock
\APACjournalVolNumPages{Transportation Research Part D: Transport and
  Environment}{95}{}{102851}.
\newblock
\begin{APACrefURL}
  [{2022-04-04}]\url{https://www.sciencedirect.com/science/article/pii/S1361920921001541}
  \end{APACrefURL}
\newblock
\begin{APACrefDOI} \doi{10.1016/j.trd.2021.102851} \end{APACrefDOI}
\PrintBackRefs{\CurrentBib}

\bibitem [\protect \citeauthoryear {%
Morelli%
\ \BBA {} Cunha%
}{%
Morelli%
\ \BBA {} Cunha%
}{%
{\protect \APACyear {2021}}%
}]{%
morelli_measuring_2021}
\APACinsertmetastar {%
morelli_measuring_2021}%
\begin{APACrefauthors}%
Morelli, A\BPBI B.%
\BCBT {}\ \BBA {} Cunha, A\BPBI L.%
\end{APACrefauthors}%
\unskip\
\newblock
\APACrefYearMonthDay{2021}{{\APACmonth{04}}}{}.
\newblock
{\BBOQ}\APACrefatitle {Measuring urban road network vulnerability to extreme
  events: {An} application for urban floods} {Measuring urban road network
  vulnerability to extreme events: {An} application for urban floods}.{\BBCQ}
\newblock
\APACjournalVolNumPages{Transportation Research Part D: Transport and
  Environment}{93}{}{102770}.
\newblock
\begin{APACrefURL}
  [{2022-04-04}]\url{https://www.sciencedirect.com/science/article/pii/S1361920921000742}
  \end{APACrefURL}
\newblock
\begin{APACrefDOI} \doi{10.1016/j.trd.2021.102770} \end{APACrefDOI}
\PrintBackRefs{\CurrentBib}

\bibitem [\protect \citeauthoryear {%
Norris%
, Stevens%
, Pfefferbaum%
, Wyche%
\BCBL {}\ \BBA {} Pfefferbaum%
}{%
Norris%
\ \protect \BOthers {.}}{%
{\protect \APACyear {2008}}%
}]{%
norris_community_2008}
\APACinsertmetastar {%
norris_community_2008}%
\begin{APACrefauthors}%
Norris, F\BPBI H.%
, Stevens, S\BPBI P.%
, Pfefferbaum, B.%
, Wyche, K\BPBI F.%
\BCBL {}\ \BBA {} Pfefferbaum, R\BPBI L.%
\end{APACrefauthors}%
\unskip\
\newblock
\APACrefYearMonthDay{2008}{{\APACmonth{03}}}{}.
\newblock
{\BBOQ}\APACrefatitle {Community {Resilience} as a {Metaphor}, {Theory}, {Set}
  of {Capacities}, and {Strategy} for {Disaster} {Readiness}} {Community
  {Resilience} as a {Metaphor}, {Theory}, {Set} of {Capacities}, and {Strategy}
  for {Disaster} {Readiness}}.{\BBCQ}
\newblock
\APACjournalVolNumPages{American Journal of Community
  Psychology}{41}{1}{127--150}.
\newblock
\begin{APACrefURL}
  [{2022-03-25}]\url{https://doi.org/10.1007/s10464-007-9156-6}
  \end{APACrefURL}
\newblock
\begin{APACrefDOI} \doi{10.1007/s10464-007-9156-6} \end{APACrefDOI}
\PrintBackRefs{\CurrentBib}

\bibitem [\protect \citeauthoryear {%
Papilloud%
\ \BBA {} Keiler%
}{%
Papilloud%
\ \BBA {} Keiler%
}{%
{\protect \APACyear {2021}}%
}]{%
papilloud_vulnerability_2021}
\APACinsertmetastar {%
papilloud_vulnerability_2021}%
\begin{APACrefauthors}%
Papilloud, T.%
\BCBT {}\ \BBA {} Keiler, M.%
\end{APACrefauthors}%
\unskip\
\newblock
\APACrefYearMonthDay{2021}{{\APACmonth{11}}}{}.
\newblock
{\BBOQ}\APACrefatitle {Vulnerability patterns of road network to extreme floods
  based on accessibility measures} {Vulnerability patterns of road network to
  extreme floods based on accessibility measures}.{\BBCQ}
\newblock
\APACjournalVolNumPages{Transportation Research Part D: Transport and
  Environment}{100}{}{103045}.
\newblock
\begin{APACrefURL}
  [{2022-04-04}]\url{https://www.sciencedirect.com/science/article/pii/S1361920921003424}
  \end{APACrefURL}
\newblock
\begin{APACrefDOI} \doi{10.1016/j.trd.2021.103045} \end{APACrefDOI}
\PrintBackRefs{\CurrentBib}

\bibitem [\protect \citeauthoryear {%
Reggiani%
}{%
Reggiani%
}{%
{\protect \APACyear {2013}}%
}]{%
reggiani_network_2013}
\APACinsertmetastar {%
reggiani_network_2013}%
\begin{APACrefauthors}%
Reggiani, A.%
\end{APACrefauthors}%
\unskip\
\newblock
\APACrefYearMonthDay{2013}{{\APACmonth{07}}}{}.
\newblock
{\BBOQ}\APACrefatitle {Network resilience for transport security: {Some}
  methodological considerations} {Network resilience for transport security:
  {Some} methodological considerations}.{\BBCQ}
\newblock
\APACjournalVolNumPages{Transport Policy}{28}{}{63--68}.
\newblock
\begin{APACrefURL}
  [{2022-03-25}]\url{https://www.sciencedirect.com/science/article/pii/S0967070X12001552}
  \end{APACrefURL}
\newblock
\begin{APACrefDOI} \doi{10.1016/j.tranpol.2012.09.007} \end{APACrefDOI}
\PrintBackRefs{\CurrentBib}

\bibitem [\protect \citeauthoryear {%
Serre%
\ \BBA {} Heinzlef%
}{%
Serre%
\ \BBA {} Heinzlef%
}{%
{\protect \APACyear {2018}}%
}]{%
serre_assessing_2018}
\APACinsertmetastar {%
serre_assessing_2018}%
\begin{APACrefauthors}%
Serre, D.%
\BCBT {}\ \BBA {} Heinzlef, C.%
\end{APACrefauthors}%
\unskip\
\newblock
\APACrefYearMonthDay{2018}{{\APACmonth{09}}}{}.
\newblock
{\BBOQ}\APACrefatitle {Assessing and mapping urban resilience to floods with
  respect to cascading effects through critical infrastructure networks}
  {Assessing and mapping urban resilience to floods with respect to cascading
  effects through critical infrastructure networks}.{\BBCQ}
\newblock
\APACjournalVolNumPages{International Journal of Disaster Risk
  Reduction}{30}{}{235--243}.
\newblock
\begin{APACrefURL}
  [{2022-03-25}]\url{https://www.sciencedirect.com/science/article/pii/S2212420918301985}
  \end{APACrefURL}
\newblock
\begin{APACrefDOI} \doi{10.1016/j.ijdrr.2018.02.018} \end{APACrefDOI}
\PrintBackRefs{\CurrentBib}

\bibitem [\protect \citeauthoryear {%
Sharma%
, Tabandeh%
\BCBL {}\ \BBA {} Gardoni%
}{%
Sharma%
\ \protect \BOthers {.}}{%
{\protect \APACyear {2018}}%
}]{%
sharma_resilience_2018}
\APACinsertmetastar {%
sharma_resilience_2018}%
\begin{APACrefauthors}%
Sharma, N.%
, Tabandeh, A.%
\BCBL {}\ \BBA {} Gardoni, P.%
\end{APACrefauthors}%
\unskip\
\newblock
\APACrefYearMonthDay{2018}{{\APACmonth{04}}}{}.
\newblock
{\BBOQ}\APACrefatitle {Resilience analysis: a mathematical formulation to model
  resilience of engineering systems} {Resilience analysis: a mathematical
  formulation to model resilience of engineering systems}.{\BBCQ}
\newblock
\APACjournalVolNumPages{Sustainable and Resilient
  Infrastructure}{3}{2}{49--67}.
\newblock
\begin{APACrefURL}
  [{2022-03-25}]\url{https://doi.org/10.1080/23789689.2017.1345257}
  \end{APACrefURL}
\newblock
\APACrefnote{Publisher: Taylor \& Francis \_eprint:
  https://doi.org/10.1080/23789689.2017.1345257}
\newblock
\begin{APACrefDOI} \doi{10.1080/23789689.2017.1345257} \end{APACrefDOI}
\PrintBackRefs{\CurrentBib}

\bibitem [\protect \citeauthoryear {%
Singh%
, Sinha%
, Vijhani%
\BCBL {}\ \BBA {} Pahuja%
}{%
Singh%
\ \protect \BOthers {.}}{%
{\protect \APACyear {2018}}%
}]{%
singh_vulnerability_2018}
\APACinsertmetastar {%
singh_vulnerability_2018}%
\begin{APACrefauthors}%
Singh, P.%
, Sinha, V\BPBI S\BPBI P.%
, Vijhani, A.%
\BCBL {}\ \BBA {} Pahuja, N.%
\end{APACrefauthors}%
\unskip\
\newblock
\APACrefYearMonthDay{2018}{{\APACmonth{06}}}{}.
\newblock
{\BBOQ}\APACrefatitle {Vulnerability assessment of urban road network from
  urban flood} {Vulnerability assessment of urban road network from urban
  flood}.{\BBCQ}
\newblock
\APACjournalVolNumPages{International Journal of Disaster Risk
  Reduction}{28}{}{237--250}.
\newblock
\begin{APACrefURL}
  [{2022-03-25}]\url{https://www.sciencedirect.com/science/article/pii/S2212420918303261}
  \end{APACrefURL}
\newblock
\begin{APACrefDOI} \doi{10.1016/j.ijdrr.2018.03.017} \end{APACrefDOI}
\PrintBackRefs{\CurrentBib}

\bibitem [\protect \citeauthoryear {%
{US Department of Homeland Security}%
}{%
{US Department of Homeland Security}%
}{%
{\protect \APACyear {2004}}%
}]{%
us_department_of_homeland_security_national_2004}
\APACinsertmetastar {%
us_department_of_homeland_security_national_2004}%
\begin{APACrefauthors}%
{US Department of Homeland Security}.%
\end{APACrefauthors}%
\unskip\
\newblock
\APACrefYear{2004}.
\newblock
\APACrefbtitle {National {Response} {Plan}} {National {Response} {Plan}}.
\newblock
\APACaddressPublisher{}{US Department of Homeland Security}.
\PrintBackRefs{\CurrentBib}

\bibitem [\protect \citeauthoryear {%
Weilant%
, Strong%
\BCBL {}\ \BBA {} Miller%
}{%
Weilant%
\ \protect \BOthers {.}}{%
{\protect \APACyear {2019}}%
}]{%
weilant_incorporating_2019}
\APACinsertmetastar {%
weilant_incorporating_2019}%
\begin{APACrefauthors}%
Weilant, S.%
, Strong, A.%
\BCBL {}\ \BBA {} Miller, B\BPBI M.%
\end{APACrefauthors}%
\unskip\
\newblock
\APACrefYearMonthDay{2019}{{\APACmonth{10}}}{}.
\newblock
\APACrefbtitle {Incorporating {Resilience} into {Transportation} {Planning} and
  {Assessment}} {Incorporating {Resilience} into {Transportation} {Planning}
  and {Assessment}}\ \APACbVolEdTR{}{\BTR{}}.
\newblock
\APACaddressInstitution{}{RAND Corporation}.
\newblock
\begin{APACrefURL}
  [{2022-03-25}]\url{https://www.rand.org/pubs/research_reports/RR3038.html}
  \end{APACrefURL}
\newblock
\begin{APACrefDOI} \doi{https://doi.org/10.7249/RR3038} \end{APACrefDOI}
\PrintBackRefs{\CurrentBib}

\end{thebibliography}





\pagebreak
\begin{appendices}
\section{Visualization Maps for the Case Study Area}
\label{sec:Visualization Maps for the Case Study Area}
\renewcommand\thefigure{\thesection\arabic{figure}}
\setcounter{figure}{0}

\begin{figure}[!h]
	\centering
    \includegraphics[width=0.9\linewidth]{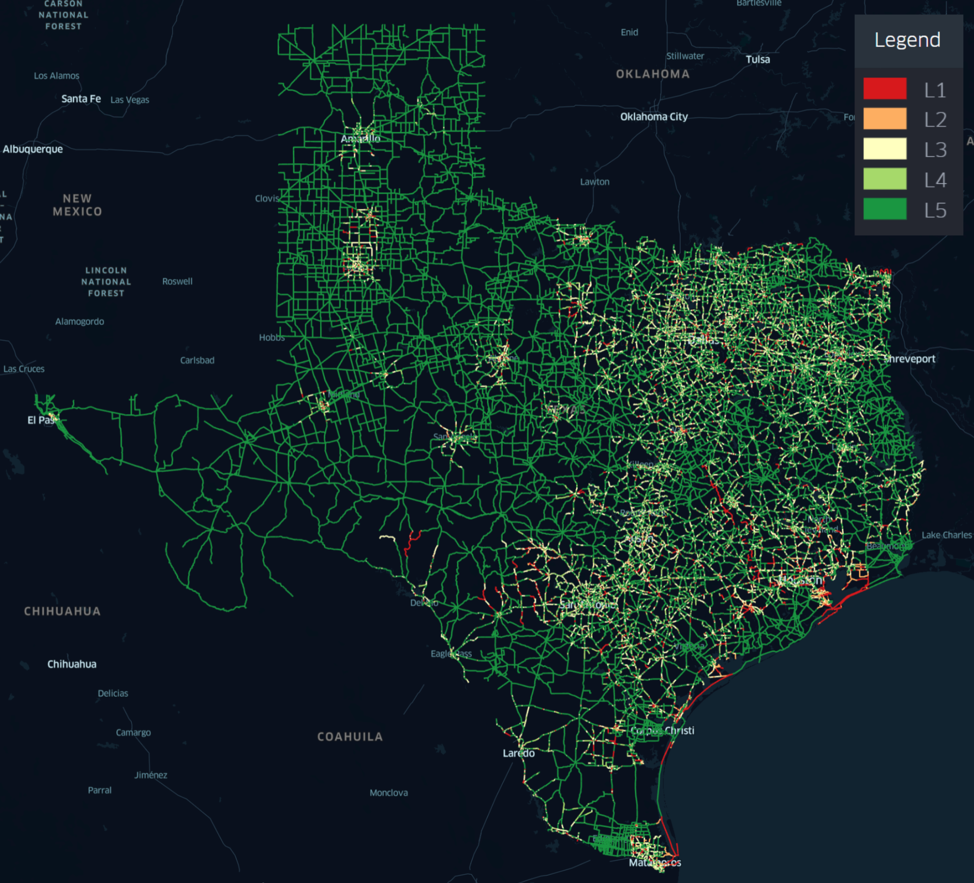}
    \caption{Criticality scores of road segments based on connectivity metrics.}
	\label{fig:figa1}
\end{figure}

\begin{figure}
	\centering
    \includegraphics[width=0.9\linewidth]{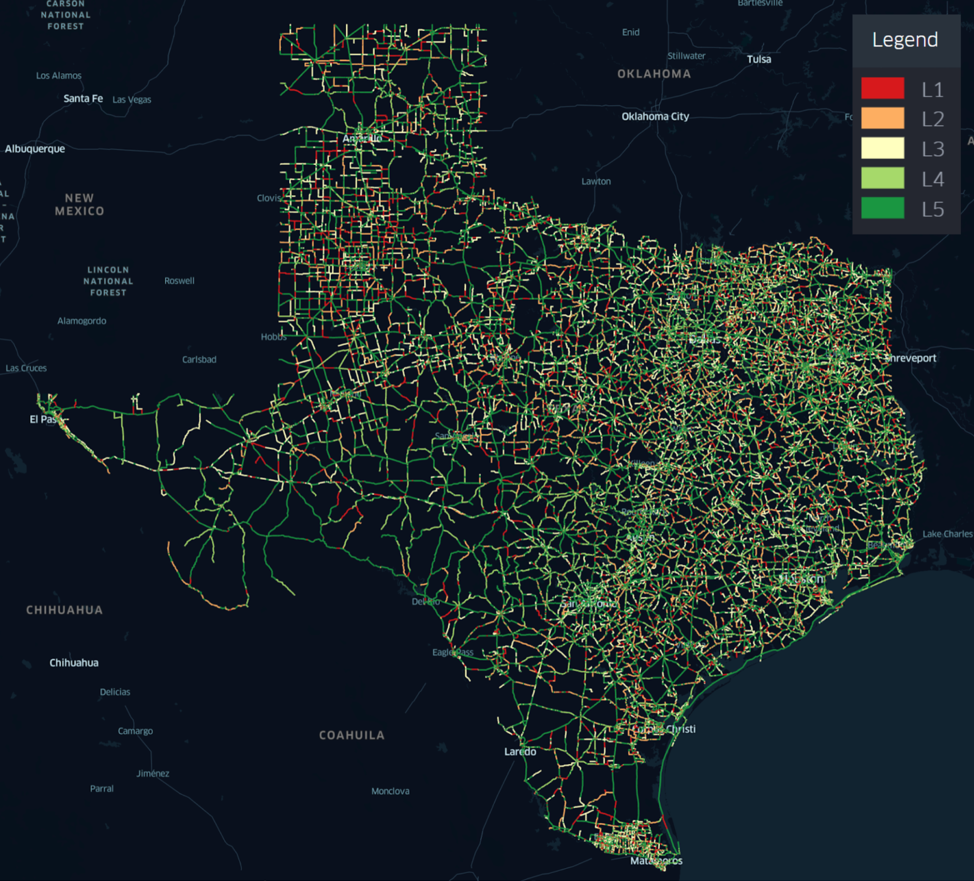}
    \caption{Criticality scores of road segments based on flood exposure.}
	\label{fig:figa2}
\end{figure}

\begin{figure}
	\centering
    \includegraphics[width=0.9\linewidth]{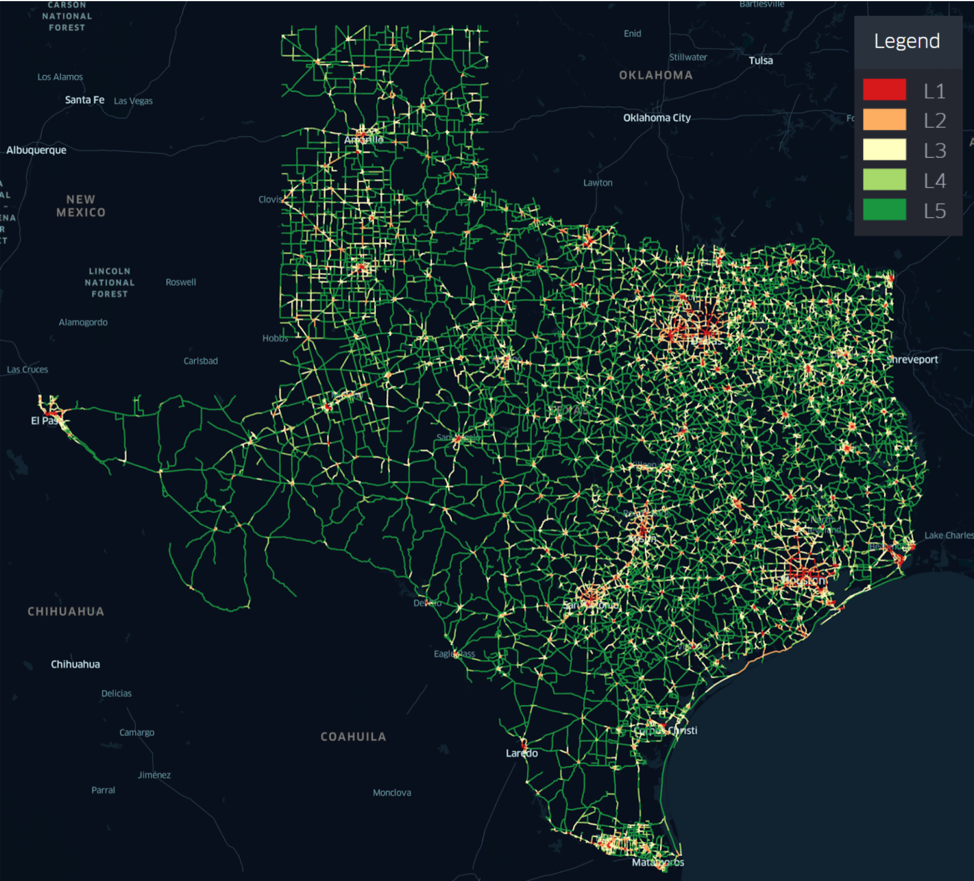}
    \caption{Criticality scores of road segments based on proximity to critical facilities.}
	\label{fig:figa3}
\end{figure}

\begin{figure}
	\centering
    \includegraphics[width=0.9\linewidth]{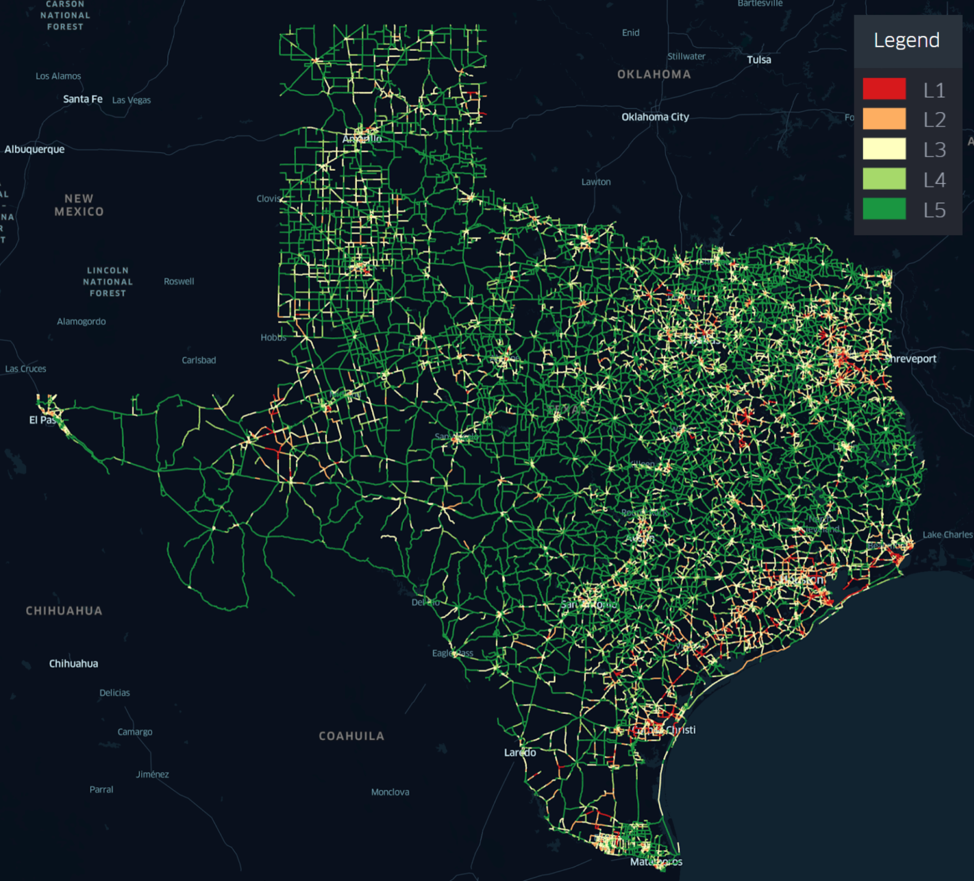}
    \caption{Criticality of road segments based on proximity to critical infrastructure networks.}
	\label{fig:figa4}
\end{figure}

\begin{figure}
	\centering
    \includegraphics[width=0.9\linewidth]{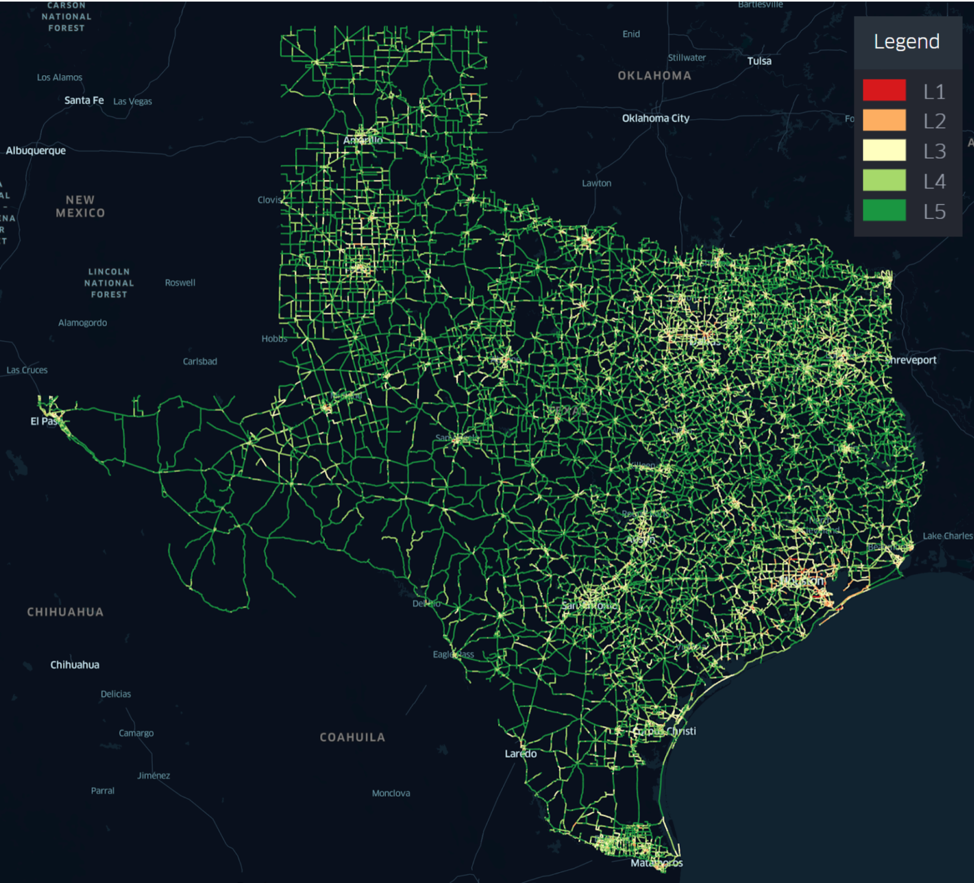}
    \caption{Integrated criticality scores based on the EWS method.}
	\label{fig:figa5}
\end{figure}

\begin{figure}
	\centering
    \includegraphics[width=0.9\linewidth]{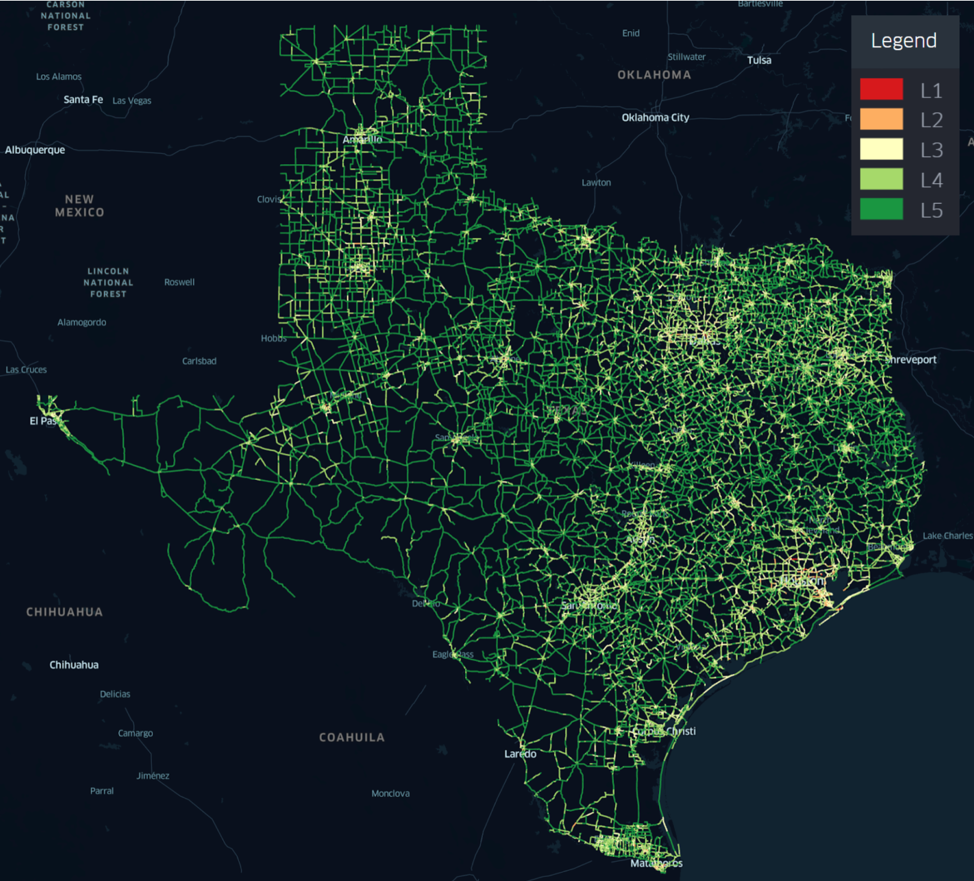}
    \caption{Integrated criticality scores based on the RSS method.}
	\label{fig:figa6}
\end{figure}

\end{appendices}

\end{document}